\documentclass[aps,prx,amsmath,amssymb,twocolumn,superscriptaddress]{revtex4-2}

\usepackage{graphicx}
\usepackage{bm}
\usepackage{hyperref}
\usepackage[capitalise]{cleveref}
\usepackage[dvipsnames]{xcolor}
\usepackage{cancel}
\usepackage[caption=false,font=normalfont]{subfig}
\usepackage{floatrow}

\usepackage{tikz}
\usetikzlibrary{positioning, shapes.geometric}
\usepackage{pdfpages}
\makeatletter
\AtBeginDocument{\let\LS@rot\@undefined}
\makeatother

\usepackage{microtype}

\usepackage{glossaries}
\glsdisablehyper
\newacronym{mps}{MPS}{matrix product states}
\newacronym{mpo}{MPO}{matrix product operator}
\newacronym{ed}{ED}{exact diagonalization}
\newacronym{tn}{TN}{tensor network}
\newacronym{tdvp}{TDVP}{time-dependent variational principle}
\newacronym{gf}{GF}{Green's function}
\newacronym{gfz}{GFZ}{Green's function zeros}
\newacronym{nnC}{nnC}{nearest-neighbour Coulomb}
\newacronym{dmrg}{DMRG}{density matrix renormalization group}
\newacronym{tknn}{TKNN}{Thouless-Kohmoto-Nightingale-Nijs}
\newacronym{dmft}{DMFT}{dynamical mean-field theory}
\newacronym{1d}{1D}{one-dimensional}
\newacronym{bbc}{BBC}{bulk-boundary correspondance}
\newacronym{sshh}{SSHH}{Su-Schrieffer-Heeger-Hubbard}
\newacronym{ssh}{SSH}{Su-Schrieffer-Heeger}
\newacronym{pbc}{PBC}{periodic boundary conditions}
\newacronym{obc}{OBC}{open boundary conditions}
\newacronym{aklt}{AKLT}{Affleck-Kennedy-Lieb-Tasaki}

\newcommand\ket[1]{\vert #1 \rangle}

\newcommand\norm[1]{\|#1\|}

\renewcommand\vec\boldsymbol

\newcommand\updt[1]{\textcolor{black}{#1}}

\begin{document}
%%%%%%%%%%%%%%%%%%%%%%%%%%%%%%%%%%%%%%%%%%%%%%%%%%%%%%%

\title{Probing Green's Function Zeros by Co-tunneling through Mott Insulators}

%%%%%%%%%%%%%%%%%%%%%%%%%%%%%%%%%%%%%%%%%%%%%%%%%%%%%%%

\newcommand\tudresdenaffil{Institute of Theoretical Physics, Technische Universit\"at Dresden}
\newcommand\ctqmataffil{W\"urzburg-Dresden Cluster of Excellence ct.qmat, 97074 W\"urzburg, Germany}
\newcommand\wuerzburgaffil{Institut f\"ur Theoretische Physik und Astrophysik,
	Universit\"at W\"urzburg, 97074 W\"urzburg, Germany
}

\newcommand\pksaffil{Max Planck Institute for the Physics of Complex Systems, N\"othnitzer Str.~38, 01187 Dresden, Germany}

\newcommand\HHaffil{I. Institute of Theoretical Physics, University of Hamburg, Notkestrasse 9, 22607 Hamburg, Germany}

\author{Carl Lehmann}
\email{Carl.Lehmann1@tu-dresden.de}
\affiliation{\tudresdenaffil}
\affiliation{\ctqmataffil}
\author{Lorenzo Crippa}
\affiliation{\wuerzburgaffil}
\affiliation{\ctqmataffil}
\affiliation{\HHaffil}
\author{Giorgio Sangiovanni}
\affiliation{\wuerzburgaffil}
\affiliation{\ctqmataffil}
\author{Jan Carl Budich}
\email{jan.budich@tu-dresden.de}
\affiliation{\tudresdenaffil}
\affiliation{\ctqmataffil}
\affiliation{\pksaffil}
\date{\today}

%%%%%%%%%%%%%%%%%%%%% ABSTRACT %%%%%%%%%%%%%%%%%%%%%%%%%%%%%%%%%%
\begin{abstract}
Quantum tunneling experiments have provided deep insights into basic excitations occurring as Green's function poles in the realm of complex quantum matter. However, strongly correlated quantum materials also allow for Green's functions zeros (GFZ) that may be seen as an antidote to the familiar poles, and have so far largely eluded direct experimental study. Here, we propose and investigate theoretically how co-tunneling through Mott insulators enables direct access to the shadow band structure of GFZ. In particular, we derive an effective Hamiltonian for the GFZ that is shown to govern the co-tunneling amplitude and reveal fingerprints of many-body correlations clearly distinguishing the GFZ structure from the underlying free Bloch band structure of the system. Our perturbative analytical results are corroborated by numerical data both in the framework of exact diagonalization and matrix product state simulations for a one-dimensional model system consisting of a Su-Schrieffer-Heeger-Hubbard model coupled to two single level quantum dots.  

\end{abstract}
%%%%%%%%%%%%%%%%%%%%%%%%%%%%%%%%%%%%%%%%%%%%%%%%%%%%%%%%%%%%%

\maketitle
%%%%%%%%%%%%%%%%%%%%%%%%%%%%%%%%%%%%%%%%%%%%%%%%%%%%%%%%%%%%%

Quantum tunneling and transport probes are among the most versatile diagnostic tools for gaining insight into the microscopic structure of quantum matter \cite{Datta_1995,Ryndyk_2009,Ihn_2009,Wiel_2002}. In the realm of strongly correlated systems \cite{Anisimov_2010,Avella_2011,Qin_2022,Arovas_2022}, experimental breakthroughs along these lines range from single electron control in co-tunneling experiments through quantum dots in the Coulomb blockade regime \cite{Barthelemy_2013,Nichol_2022,Hensgens_2017,Qiao_2020,Pustilnik_2004,Wang_2023} to probing the fractional statistics of anyonic quasiparticles in topological phases \cite{Banerjee_2018,Kasahara_2018,Bartolomei_2020,Nakamura_2020}. However, despite this remarkable progress, the complexity of correlated many-body systems still retains a plethora of secrets, thus promising novel physical phenomena yet to be revealed. As an exciting example, \gls{gfz} in Mott insulators \cite{Dzyaloshinskii_2003,Stanescu_2007,Sakai_2009,Sakai_2010,Dave_2013,Rosch_2007} have recently been predicted to appear as a shadow of the underlying Bloch band structure~\cite{Gurarie_2011,Gurarie_2011_2,Bollmann_2024,Zhao_2023,Setty_2024,Gavensky_2023,Pasqua_2024,Blason_2023,Wagner_2023,Wagner_2024}, i.e., of the poles in the single particle \gls{gf} of a quantum material. Owing to Cauchy's argument principle, such \gls{gfz} and poles play a reciprocal role reminiscent of matter and anti-matter in particle physics, and their cancellation at material interfaces has been recently discussed in the context of topological edge modes \cite{Wagner_2023}. 
Despite recent progress regarding the possibility of detecting GFZ either by exploiting zero-pole annihilation mechanisms \cite{Wagner_2023,Wagner_2024,merinoArXiV,pepinArXiV} or through transport signatures \cite{Bollmann_2024}, it is fair to say that a direct experimental probe of GFZ has largely remained elusive.

\begin{figure}[htp]
	\centering
	\includegraphics[width=1.0\linewidth]{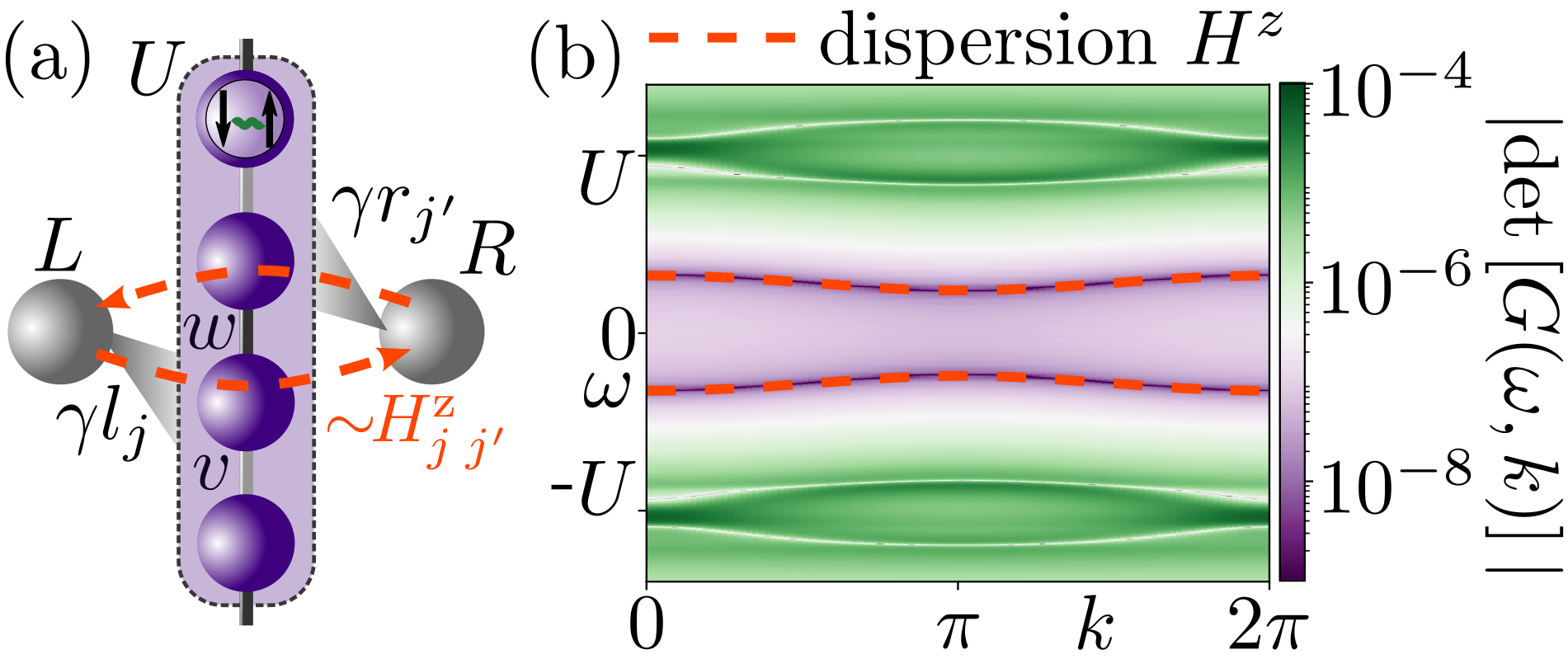}
	\caption{Co-tunneling reveals \gls{gfz}-Hamiltonian $H^{\text{z}}$: (a) Two spinful auxiliary sites (grey, $\{L,R\}$), weakly coupled by tunneling amplitudes $\{\gamma l_j,\gamma r_{j'} \}$ to a Hubbard system in Mott regime (violet). Elastic co-tunneling establishes an effective coupling between $L$ and $R$ (red, dashed line).
		(b) Numerical results (\gls{mps}) on the \gls{gf}-determinant $\vert \text{det}[G(\omega,k)]\vert$: SSH-Hubbard model at half-filling with $N=60$ spinful sites, $v=0.75,w=1.0$, $U=10$ and analytic prediction of the in-gap zero dispersion of $H^{\text{z}}$ (red, dashed line).}
	\label{fig:pole_zero_intro}
\end{figure}

Here, we study co-tunneling through Mott insulators as a sharp diagnostic tool to fully reveal the shadow band structure of \gls{gfz}. 
As a setting (see Fig.~\ref{fig:pole_zero_intro} for an illustration), we propose a re-interpretation and generalization of the well-established experimental platform of co-tunneling measurements through quantum dots \cite{Barthelemy_2013,Schleser_2005,Braakman_2013,Sanchez_2014,Bordin_2024,Wang_2023,Ranni_2021,Schindele_2014,Liu_2022} by replacing a quantum dot in the Coulomb-blockade regime with a many-body system in a Mott phase, which may even be engineered as a synthetic quantum material consisting of an array of coupled quantum dots \cite{Nichol_2022,Barthelemy_2013,Hensgens_2017,Bayer_2024}. 
We pinpoint two fundamental differences between our strongly correlated scenario and a tunneling probe for an uncorrelated band insulator. 
First, the band structure of \gls{gfz} appears in the numerator of the co-tunneling amplitude, contrary to an energy denominator for an uncorrelated insulator.
Second, genuine correlation effects decisively modify the zero-bands compared to the underlying free band structure of the Mott insulator. Using the striking example of topological distinction between zero-bands and the corresponding pole-bands, we demonstrate how these subtle fingerprints of many-body correlations can be directly observed through the co-tunneling amplitude in mesoscopic quantum transport experiments.

\paragraph*{Zeros band structure through perturbation theory} --
To describe how to probe a solid-state system spectroscopically, we use the formalism of Green's functions. At the one-particle level, this allows to quantify the response of the system upon adding or removing an electron, taking into account correlation effects.
The single-particle electronic Green's function in the K\"all{\'e}n-Lehmann representation at zero temperature is explicitly dependent on the spectrum of excitations of the system~\cite{Mahan_1990,Bruus_2004}:
% \begin{align}
% 	G_{j,\sigma;j',\sigma'}(\omega) &= \sum_{\nu} \Bigg[\frac{\langle GS \vert c_{j,\sigma}\vert E_\nu \rangle \langle E_\nu \vert c_{j',\sigma'}^\dagger \vert GS \rangle }{\omega + E_{GS} - E_\nu} \nonumber \\
% 	&+ \frac{ \langle GS \vert c_{j',\sigma'}^\dagger \vert E_\nu \rangle \langle E_\nu \vert c_{j,\sigma} \vert  GS \rangle }{\omega + E_\nu - E_{GS}} \Bigg] \text{.}
% 	\label{EQ:kl_green}
% \end{align}
\updt{\begin{align}
	G_{j,\sigma;j',\sigma'}(\omega) &= \Bigl< GS \Big| \Bigl[ c_{j,\sigma} (\omega + E_{GS} - H_{\text{H}} )^{-1} c_{j',\sigma'}^\dagger  \nonumber \\ &~~ +  c_{j',\sigma'}^\dagger (\omega - E_{GS} + H_{\text{H}}  )^{-1}  c_{j,\sigma} \Bigr] \Big| GS \Bigr>
     \text{,}
	\label{EQ:kl_green}
\end{align}}
where $c^{(\dagger)}_{j,\sigma}$ denotes the annihilation and creation operators for site $j$ and spin $\sigma$. In the non-interacting limit, the Hamiltonian and $G(\omega)$ are readily diagonalizable in a single-particle basis. The determination of the spectral and geometrical properties of the system then amounts to the study of the poles of $G(\omega)$, each of which represents an infinitely long-lived single-particle eigenmode.
However, in presence of correlations, a large amount of amplitudes in the numerator become nonzero, reflecting the multitude of scattering events promoted by the many-body interaction term. As a result, the spectral weight distribution is broadened in the frequency-momentum plane. There is, however, a remarkable consequence of this scenario: in the spectral distribution, a set of isolated Green's function zeros (\gls{gfz}) can emerge. Their presence has been linked to peculiar thermodynamic \cite{Fabrizio_2022} and topological properties \cite{Gurarie_2011,Gurarie_2011_2,Wagner_2023,Wagner_2024,Bollmann_2024,Zhao_2023,Setty_2024,Gavensky_2023,Pasqua_2024,Blason_2023}, and the study of their dispersion is  therefore an intriguing, if nontrivial endeavor.

The \gls{gfz} are defined as zero eigenvalues of $G$, or equivalently as the points at which its determinant vanishes~\cite{Gurarie_2011}. 
In the noninteracting limit, there are no isolated zeros. In the opposite case, where the system is deep in the Mott phase, two incoherent Hubbard bands form at high frequency, which contain the totality of the poles of the Green's function, and a large number of zeros. However, inside the hard Mott gap between the two Hubbard bands, no poles are present. By contrast, GFZ can, and do, manifest at low energy, irrespective of the system size.  The focus of this work lies on how to detect these isolated in-gap \gls{gfz}, which are entirely responsible for the low energy properties of the \gls{gf} of Mott insulators~\cite{Dzyaloshinskii_2003,Stanescu_2007,Sakai_2009,Sakai_2010,Dave_2013,Rosch_2007,Gurarie_2011,Gurarie_2011_2,Bollmann_2024,Zhao_2023,Setty_2024,Gavensky_2023,Pasqua_2024,Blason_2023,Wagner_2023,Wagner_2024}.

As a first step in our discussion, we derive a remarkably simple and intuitive expression for $G$ at low energies ($\omega \ll U$) \updt{deep in the Mott regime at half-filling, and with a paramagnetic groundstate}.  
Our non-interacting Hamiltonian $H_0=\sum h_{j;j'}c^{\dagger}_{j,\sigma}c_{j',\sigma}$ consists of a linear combination of on-site terms, which we denote by $M$, and non-local hopping terms ($W$).
The interaction term is of the Hubbard type $UH_{\text{int}} = 2U\sum_{j} (n_{j,\downarrow} - \frac{1}{2})(n_{j,\uparrow} - \frac{1}{2})$, where $j$ labels the lattice site and $\sigma=\uparrow,\downarrow$ is the spin. The two competing energy scales are the bandwidth of $H_{0}$, $2D$, and the local Coulomb interaction strength $U$, where $U\gg D$ for a Mott insulator. For convenience, we introduce the rescaled non-interacting Hamiltonian term $H^{\prime}_0 = \frac{1}{D} H_0$, as well as the small parameter $\phi = \frac{D}{U}$. The Hamiltonian can therefore be rewritten as
\begin{equation}
    H_{\text{H}} = U(\phi H^{\prime}_{0} + H_{\text{int}}).
    \label{EQ:general_H}
\end{equation}
In this form, $H_{\text{H}}$ naturally suggests a perturbative analysis in $\phi$: indeed, by applying degenerate perturbation theory~\cite{Sakurai_2020} up to first order in $\phi$ we determine its perturbed eigenenergies $E_\nu^P$ and eigenstates  $\vert E_\nu^P \rangle $, which we can then conveniently insert in Eq.~\eqref{EQ:kl_green} to obtain the perturbed Mott Green's function. It is finally a matter of collecting all the lowest terms in $\phi$ \updt{and expanding the denominator at low energies}, to arrive at a key expression for the Green's function, which is valid for \updt{ $\phi \ll 1$ and $\omega\ll U$ \footnote{Due to expansions of $\frac{1}{\omega-\omega_0}$ terms, our approximation breaks down at large energies $\omega\approx U$.}}: 
\begin{align}
   G_{j,\sigma;j',\sigma'}(\omega)\,\, {\approx} \, \,\frac{1}{U^2} \left[- \omega \mathbb{I} + H^{\text{z}} \right]_{j,\sigma;j',\sigma'} + \mathcal{O}(\phi^3).
   \label{EQ:gf_hubbard_intro}
\end{align}
This expression is intuitively consistent with the strongly suppressed in-gap spectral distribution of a Mott insulator. More than that, it immediately and clearly highlights the presence, inside of the Mott gap, of points where the determinant of the Green's function is exactly zero. These are the roots of the expression $H^{z}-\omega\mathbb{I}$. In other words, and inline with \cite{Wagner_2023}, we find that the low-energy isolated zeros of a Mott insulator are described by an effective Hamiltonian $H^{\text{z}}$. The detailed derivation of Eq.~\eqref{EQ:gf_hubbard_intro} can be found in the supplemental material~\cite{reference_sm}\nocite{Zwiebach_2018,Ripoll_2022,Affleck_1987,Amaricci_2017}, \updt{including a general discussion of $SU(2)$-breaking bare Hamiltonians and groundstates}.

Our approach is able to precisely characterize the effective Hamiltonian, and leads to a second interesting result about the~\gls{gfz}: in this limit $H^{z}$ has, modulo some exactly quantifiable renormalization, the same form as the non-interacting term $H_{0}$ of the original Hamiltonian Eq.~\eqref{EQ:general_H}.

For strong Coulomb repulsion $U$ and at half-filling, the 
Hubbard model at low energies is well described by an effective spin-$1\slash 2$ model~\cite{Qin_2022,Arovas_2022,Anisimov_2010,Avella_2011}.
We can therefore rewrite the lowest-order terms in $G$ in the spin basis, thereby obtaining an expression linking the approximated Green's function to spin-spin correlators, as derived in detail in~\cite{reference_sm}. For the bare Hamiltonian $h_{j;j'}= (M+W)_{jj'}$, we find a remarkably elegant expression for the effective Hamiltonian of the~\gls{gfz}:
\begin{align}
	H^{\text{z}}_{jj'}=  M_{jj'} \delta_{j,j'}+ 4\langle   \overrightarrow{S}_j \cdot \overrightarrow{S}_{j'}  \rangle_{j\neq j'} W_{jj'},
	\label{EQ:simple_HH}
\end{align}
where $j,j'$ are lattice sites. Here, expectation values are evaluated with respect to the groundstate eigenvector $\ket{GS^{0}}$ of the corresponding effective spin-$1\slash2$ model~\footnote{That is the correct groundstate towards the unperturbed limit and follows from degenerated perturbation theory.}, and are hence accessible for a large variety of Hubbard models. The effective zeros Hamiltonian can then be decomposed into two terms, in a striking analogy with the on-site and hopping terms of the bare Hamiltonian $H_{0}$: (i) the on-site terms of the non-interacting Hamiltonian, described by $M$, are carried over in the dispersion of the zeros without renormalization, while (ii) the  hopping terms, described by $W$, are re-scaled by the spin correlation function $\langle   \overrightarrow{S}_j \cdot\overrightarrow{S}_{j'} \rangle$~\footnote{The correlation value is evaluated in the limit towards, not at, the atomic limit, making it generally non-trivial.}.  
This follows naturally from the combination of our perturbative approach around the atomic limit, and the nature of competing energy scales in the system: any quadratic on-site term commutes with $H$ in the atomic limit, and the perturbed energy is just shifted by $M$ without renormalization. On the contrary, hopping and interaction strength are in direct competition: treating the first order perturbatively results in a lowest-order correction to the atomic limit Hamiltonian, which famously depends on the spin-spin correlation function, as in the derivation of the superexchange~\cite{Anderson_1963,Koch_2017,Pavarini_2015,Arovas_2022}. \updt{These antiferromagnetic correlations characterize paramagnetic Mott insulators beyond mean field and are responsible for the sign flip in the nearest-neighbour hopping in Eq.~\eqref{EQ:simple_HH}.}

\paragraph*{Probing the zeros}--
As clearly evidenced by Eqs.~\eqref{EQ:gf_hubbard_intro} and~\eqref{EQ:simple_HH}, the knowledge of the zeros dispersion is sufficient to determine the spectral properties of the Mott system at low frequency. 
Obtaining a clear hint of the zeros behavior from experimental observations is instead less straightforward since any spectral measurement (e.g. via photoemission) would just confirm the presence of a gap, without being able to resolve the elusive zeros within.
We propose a novel approach to the problem, which is, at the same time, surprisingly simple to implement: instead of directly performing spectroscopic measurements on the Mott insulator, we employ it as the potential barrier in a co-tunneling experiment, and study the effective coupling between the leads. 
We illustrate the setup via a minimal example, which is a convenient adaption of already existing quantum dot experimental setups~\cite{Braakman_2013,Hensgens_2017,Barthelemy_2013,Nichol_2022}. This is sketched in Fig.\ref{fig:pole_zero_intro} (a) and described by $H_{\text{tot}} = H_{\text{H}} + \gamma C + H_{\text{D}}$. The system consists of two quantum dots, henceforth L and R, and a central tunneling medium. The Hamiltonian of the quantum dots, $H_{\text{D}}$, is that of a Hubbard atom with Coulomb repulsion $2U$. \updt{ In order to reveal fingerprints of the GFZ,} the central element is a spin-isotropic Hubbard model in the Mott phase, described by the previously discussed Eq.~\eqref{EQ:general_H}. The chemical potential of the quantum dots is increased by $U$ in relation to the central Hubbard model, and independently adjustable by $\mu_{\text{L}},\mu_{\text{R}}$. The system-dot coupling $C$ is given by arbitrary hopping amplitudes $\{l_i\}$ and $\{r_i\}$ between the $i$-th site of the central Mott insulator and the L and R quantum dots. For simplicity, we will, from now on, omit the spin index $\sigma$, and restrict ourselves to the case where $l_{i}=\delta_{i,j}$ and $r_{i}=\delta_{i,j'}$ for two different sites $j \neq j'$. The full derivation for the general case with arbitrary hopping array is provided in~\cite{reference_sm}.

In a co-tunneling experiment, only the leads are directly probed. This amounts to tracing out the degrees of freedom associated to the central system, and only considering the overall propagation amplitude of excitation leaving (entering) the two leads via the quantum dots. These processes are, in our proposed setting, exactly described by the one-particle Green's function of the \updt{central tunneling medium.}
In short, by applying a Schrieffer-Wolff transformation \cite{Schrieffer_1966,Bravyi_2011,Bir_1974} in the \updt{limit of weak system-dot coupling $\gamma$ with small dot potentials $\mu_{\text{L} \slash\text{R}}$ \footnote{More precise: the system-dot coupling $\gamma$ and the dot potentials $\mu_{\text{L} \slash\text{R}}$ need to be smaller than the first excitation gap above the groundstate of the central barrier. }}, and assuming a large lead-dot interaction $\sim U$, we derive an effective spin-isotropic Hamiltonian $\bar{h}$ \updt{in the low-energy subspace}. It describes the dynamics of a single particle co-tunneling between the left and right quantum dots through the \updt{central medium}:
\begin{equation}
	\bar{h} =\begin{pmatrix}
    \mu_{\text{L}} + u_{\text{L}}(\mu_{\text{L}},\mu_{\text{R}})  & t^\ast_{\text{co}}(\bar{\mu}) \\
    t_{\text{co}}(\bar{\mu})  & \mu_{\text{R}} + u_{\text{R}}(\mu_{\text{L}},\mu_{\text{R}})
    \end{pmatrix}
    \label{EQ:heff-dots}
\end{equation}
where $\bar{\mu}$ is an "average" chemical potential $\frac{1}{2} \left(\mu_{\text{L}} + \mu_{\text{R}}\right)$. The presence of the \updt{central tunneling medium} gives rise to a renormalization to the on-site energies, given by $u_{L/R}\sim\mathcal{O}(\gamma^2)$, which can be conveniently reabsorbed by tuning the quantum dot chemical potentials. The most relevant quantity in Eq.~\eqref{EQ:heff-dots} is the \updt{co-tunneling amplitude 
\begin{align}
	t_{\text{co}}(\bar{\mu},l_j,r_{j'}) = \gamma^2 G_{j,j'}(\bar{\mu}),
 \label{EQ:eff_ham_bath}
\end{align}
which describes the coupling between the left and right quantum dot mediated via the central barrier (see Fig.~\ref{fig:pole_zero_intro} for an illustration). Since we are interested in a Mott insulator at half-filling as the tunneling barrier, we can make use of the previously derived approximation for the Green's function in Eq.~\eqref{EQ:gf_hubbard_intro}.} We immediately notice how, through Eq.~\eqref{EQ:gf_hubbard_intro}, the co-tunneling amplitude $t_{\text{co}}$ is directly related to the \gls{gfz}-Hamiltonian $H^{\text{z}}$ of \updt{ Eq.~\eqref{EQ:simple_HH}. Explicitly, for $l_j$ and $r_{j'}$ (cf.~Fig.~\ref{fig:pole_zero_intro}) chosen such that $j \neq j'$ (see \cite{reference_sm} for general tunnel-couplings $l,r$):
\begin{align}
t_{\text{co}} = \frac{\gamma^2}{U^2} \left(H^{\text{z}}\right)_{j,j'}
 \label{EQ:central_result}
\end{align}
As a central result, Eq.~\ref{EQ:central_result} clarifies the correspondence between the Mott Green's function and the co-tunneling amplitude:} instead of directly probing the features of the vanishing spectral function inside the Mott gap, a nearly impossible task, we are now recasting the same information into what is effectively a potential barrier for the evanescent wave of the co-tunneling experiment. Its features, and in particular the spectroscopically undetectable zeros, have a macroscopic effect on the measured occupations of the leads.
As a further benefit, we can now take advantage of the well-established theoretical tools and experimental techniques employed in the study of quantum tunneling, such as charge stability diagrams~\cite{DiCarlo_2004,Gaudreau_2006,Wang_2011,Hensgens_2017,Foulk_2024} and time-resolved charge detection of single tunneling events \cite{Vandersypen_2004,Lu_2003,Schleser_2004,Gustavsson_2006,Gustavsson_2008,Gustavsson_2009,Braakman_2013}.

\paragraph*{Numerical results}-- We now provide numerical evidence of our key findings by considering the paradigmatic case of the one-dimensional \gls{ssh}-Hubbard model\cite{Wagner_2023,Su_1980}, consisting of a bipartite chain with alternating hoppings $v$ and $w$. We numerically solve the Hamiltonian of the co-tunneling setup by using Density Matrix Renormalization Group (DMRG)~\cite{Schollwock_2005, Schollwock_2011} in combination with the \gls{tdvp}~\cite{Yang_2020,Paeckel_2019,Goto_2019} and \gls{ed} methods. We first compute the real-time \gls{gf} and then transform it into the frequency domain, thereby gaining access to the \gls{gfz}. 

We first prove that our analytical formula for the zeros Hamiltonian is an adequate approximation of the in-gap spectral features of a Mott insulator. Fig.\ref{fig:pole_zero_intro} (b) shows a comparison between the determinant of the Green's function of the SSH-Hubbard model, obtained via DMRG, and the approximated formula for the \gls{gfz} Hamiltonian of Eq. \eqref{EQ:simple_HH}. We note that the isolated zeros (dark purple regions of the color plot), which appear in the Mott gap, are in perfect agreement with the eigenvalues of $H^{\text{z}}$, already at the intermediate value of $\phi \approx 1/10$. \updt{In the supplemental material~\cite{reference_sm}, we provide a more detailed quantitative analysis. }

\begin{figure}[htp]
	\centering
	\includegraphics[width=1.0\linewidth]{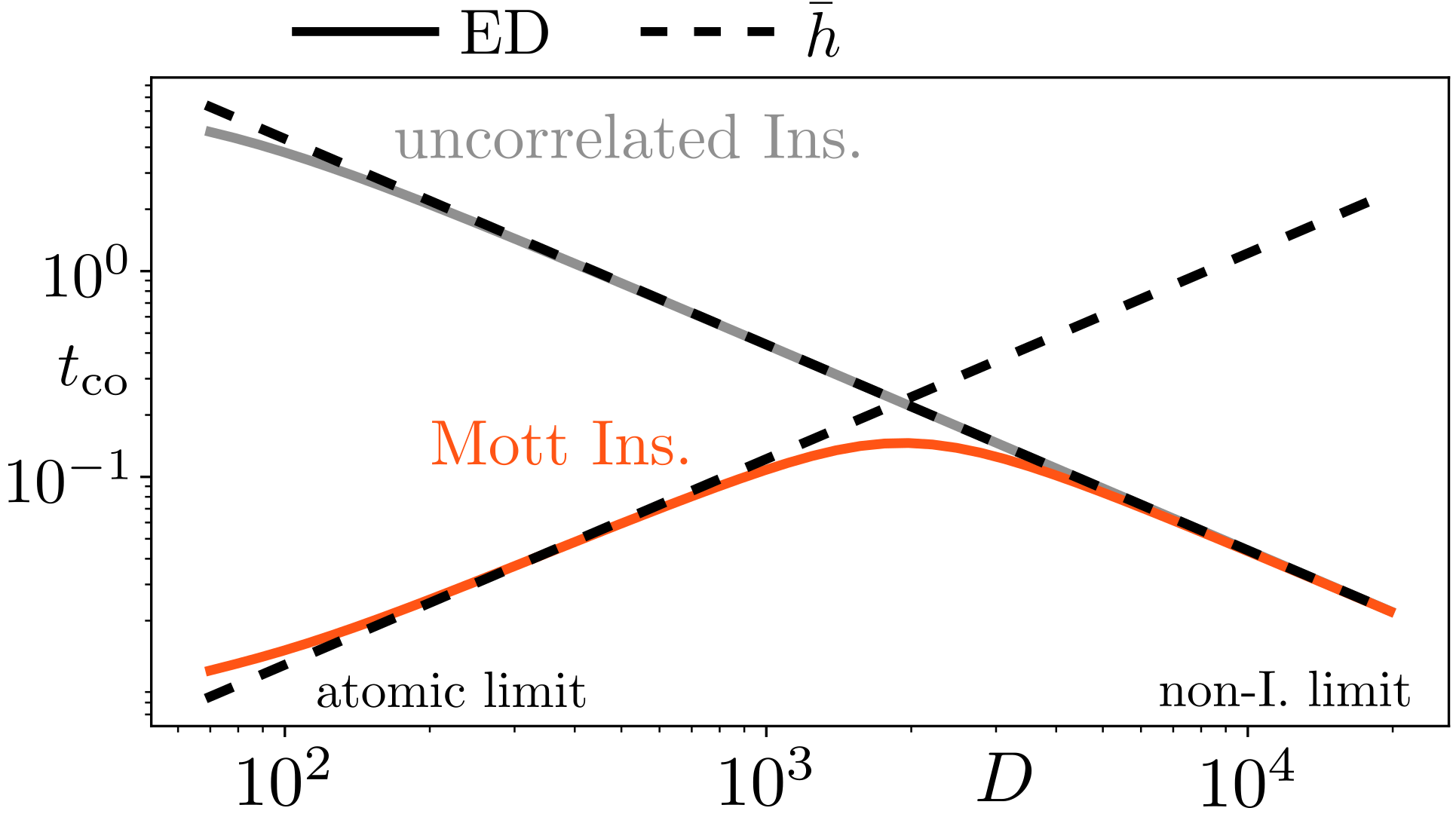}
	\caption{Co-tunneling amplitude $t_{\text{co}}$ as function of bandwidth $D$ (fixed $U =2000$, $N=16$, $\gamma=20$,$l_i = \delta_{i,2} $,$r_i = \delta_{i,3} $, $v = 0.9 w$, PBC). \gls{ssh}-Hubbard model (solid,red) shows linear $D$ dependence within the the Mott regime $\frac{D}{U} < 1$, due to energy nominator in \gls{gf} Eq.~\eqref{EQ:gf_hubbard_intro}. Non-interacting insulator (solid,grey) with energy denominator (parameters as above but $U=0$). Analytical results, Eq.\eqref{EQ:eff_ham_bath} (dashed).}
	\label{fig:den_vs_nom}
\end{figure}

We then turn our attention to the co-tunneling setup, numerically simulated by supplementing the quantum dots and coupling terms, $H_{\text{D}}$ and $C$, to the SSH-Hubbard Hamiltonian. 

In Fig.\ref{fig:den_vs_nom} we show the emergent co-tunneling amplitude $t_{\text{co}}$ for the experimental setup in Fig.\ref{fig:pole_zero_intro}(a) between two interacting auxiliary quantum dots, as a function of the bandwidth $D$. We compare its behavior in the case of a Mott insulator and a conventional non-interacting insulator, and plot the two resulting curves (red and grey) as a function of $D$. The numerically obtained curves are to be compared with the two dashed lines, which show the two different behaviors in the non-interacting and Mott limits, and can be extracted from Eq.~\eqref{EQ:eff_ham_bath}. In the non-interacting limit, the behavior of the Green's function $G$, as defined in Eq.~\eqref{EQ:kl_green}, is determined by a single energy scale, the non-interacting bandwidth appearing at the denominator~\cite{Mahan_1990,Bruus_2004}. Hence, $t_{\text{co}}\sim D^{-1}$, and the two solid lines are on top of each other and agree with the non-interacting behavior (negative slope in the logarithmic plot). At around $\frac{D}{U}\approx 1$, i.e., near the Mott transition, the behavior of the interacting system changes: now, the dominant energy scale of the system is the Coulomb interaction $U$, and approximation Eq.~\eqref{EQ:gf_hubbard_intro} holds. $H^{z}$, though renormalized as in Eq.~\eqref{EQ:simple_HH}, has a $\sim D$ dependence on the bandwidth, and the Mott co-tunneling amplitude indeed agrees with the positive-slope dashed line. The general agreement between the approximated value of $t_{\text{co}}$ and the numerically determined ones is remarkable, confirming the robustness of Eq.~\eqref{EQ:eff_ham_bath}. Only at small bandwidth $D$, where $\gamma \approx D$, does our co-tunneling approximation break down and higher-order tunneling processes become relevant.

\begin{figure}[htp]
	\centering
	\includegraphics[width=1.0\linewidth]{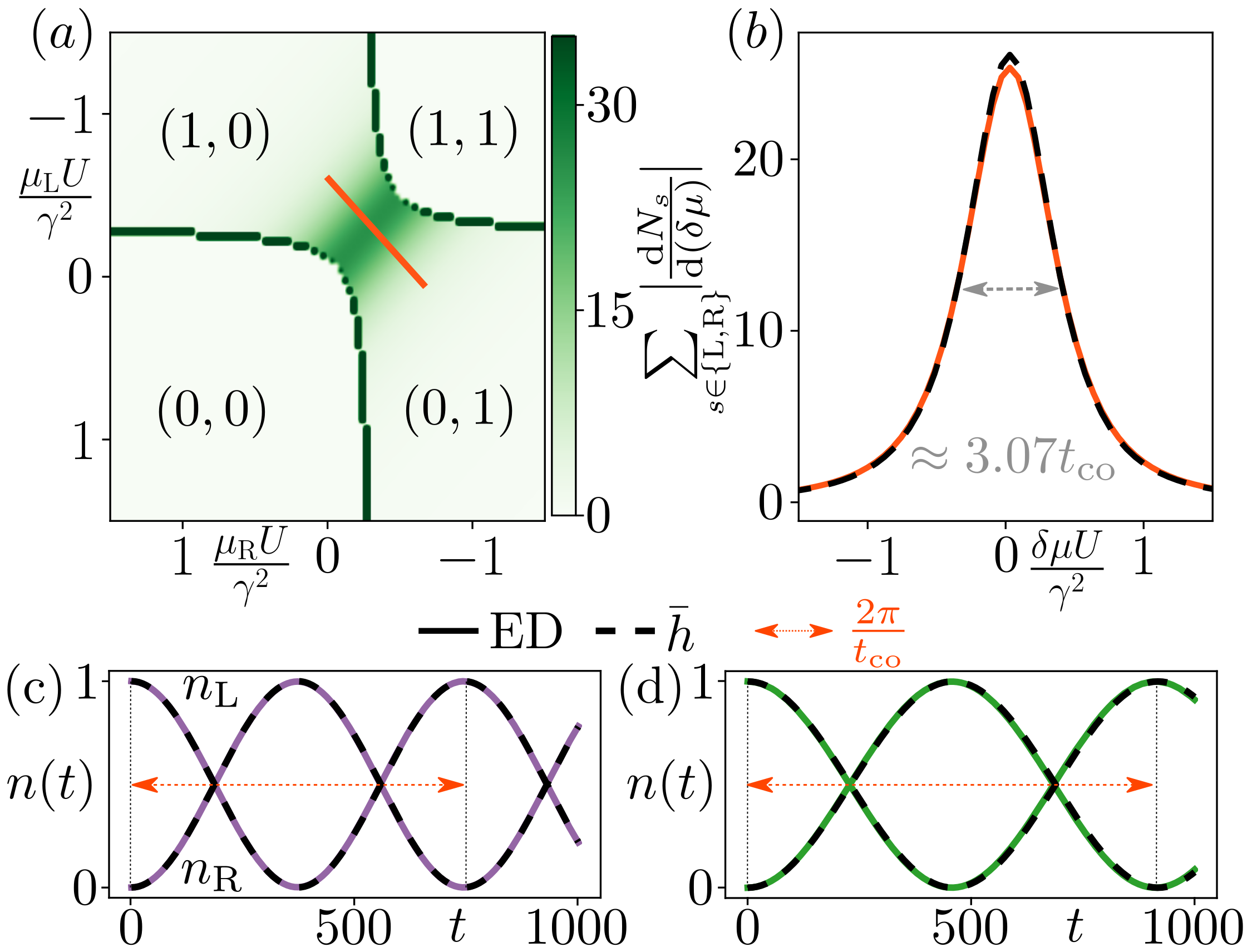}
	\caption{Charge stability diagram (\gls{ed}, $U=200$ $N=20$, $\gamma=4$,$l_i = \delta_{i,2} $,$r_i = \delta_{i,3} $, $v = 18$, $w = 20$,PBC) : (a) Derivative of the particle number $N_L$($N_R$) on left(right) dot for varying potentials $\mu_{\text{L}},\mu_{\text{R}}$. In (b): cut through (a) (red line), FWHM is proportional to co-tunneling $t_{\text{co}}$. (c) and (d): charge oscillations between aux. dots (initial localization left dot) same parameter as (a), but OBC,  $l_i = \delta_{i,1} $,$r_i = \delta_{i,2} $ in (c) and $l_i = \delta_{i,3} $,$r_i = \delta_{i,4} $ in (d). Analytical results (dashed) from Eq.~\eqref{EQ:eff_ham_bath}.}
	\label{fig:charge_stability}
\end{figure}

Our co-tunneling approach to detecting Green's function zeros can be nicely illustrated with the well-established tool of charge-stability diagrams~\cite{Hofmann_1995,DiCarlo_2004,Gaudreau_2006,Wang_2011,Hensgens_2017,Foulk_2024}, which is commonly used to describe the behavior of quantum dots as a function of control voltages. In Fig.\ref{fig:charge_stability} (a), we show the charge-stability diagram of our co-tunneling setup, by plotting the derivative of the charge occupation $N_{L\slash R}$ with respect to the detuning $\delta \mu = \mu_{\text{L}} - \mu_{\text{R}}$, at varying chemical potentials $\mu_{\text{L}},\mu_{\text{R}}$. Due to the finite tunneling amplitude $t_{\text{co}}$, the boundaries of the characteristic checkerboard structure of the uncoupled system ($\gamma=0$) are smeared out, and the localization of a single particle can be tuned smoothly from the left dot $(1,0)$ to the right dot $(0,1)$. Fig.\ref{fig:charge_stability} (b) shows a diagonal cut through Fig.\ref{fig:charge_stability} (a) at a fixed average potential $\bar{\mu} = \frac{1}{2}\left(\mu_{\text{L}} + \mu_{\text{R}}\right)$ (red line). \updt{At zero temperature}, the FWHM of the Lorentzian dispersion is directly proportional to the co-tunneling amplitude. 
An alternative way of probing the co-tunneling amplitude is to measure the frequency of Rabi-charge oscillations between the dots \cite{Schleser_2004,Gustavsson_2008,Gustavsson_2009,Braakman_2013}. It can be extracted by the time-evolution of a localized state and compared with the theoretical value $\Omega \approx \sqrt{\vert t_{\text{co}}\vert ^{2}+\omega^{2}_D} $, where $\omega_D \approx \frac{1}{2}\delta \mu  +  \mathcal{O}(\frac{\gamma^2}{U}) $ is a shifted detuning between the two quantum dots, which is easily controllable via $\mu_{L}$ and $\mu_{R}$. In Fig.\ref{fig:charge_stability} (c) and (d), we plot the numerically calculated time-dependent charge occupation values of the two quantum dots on resonance $\omega_D =0 $ (at peak maximum in~Fig.\ref{fig:charge_stability} (b)), when the system is initialized in the completely localized (1,0)-state. The data are in perfect agreement with the predicted oscillation frequency, which can be easily extracted by Eq.\eqref{EQ:heff-dots}.

\updt{The proposed probing scheme applies to generic Mott insulators, and so far our discussion has been generally applicable. As a particularly promising experimental realization, we now elaborate on a specific Hubbard system synthetically realizing a Mott insulator with coupled quantum dots.} 
In a real experiment \updt{on quantum dot arrays}, the perturbative limit $\frac{\gamma}{D}\ll \frac{D}{U} \ll 1$, may not be perfectly accessible and is further inherently tied to a small co-tunneling amplitude, making it experimentally challenging. Despite this limitation, macroscopic values of $t_{\text{co}}$ ($\geq 0.5 \mu eV$) can still be achieved for a broad range of typical parameters $U,D$ and $\gamma$, within an acceptable quantitative deviation ($\leq 10 \%$, see \cite{reference_sm} for details). Interestingly, we also find that the striking \updt{re-scaling in Eq.\eqref{EQ:simple_HH} by spin-spin correlations} can be probed qualitatively, without requiring fine-tuning of the dots. More specifically, we find a notable amplification of the tunneling amplitude as function of varying boundary conditions, a characteristic signature of this renormalization (see \cite{reference_sm} for further details). 

\noindent

\paragraph*{Conclusion}-- Starting from a degenerate perturbation theory around the atomic limit, we have derived a simple yet highly accurate approximation for the low-frequency features of a Mott Green's function, which can be completely described by a dispersion relation for the Green's function zeros. This situation is reminiscent of a non-interacting system, with the important distinction that the effective dispersion of zeros appears in the numerator of the Green's functions, and is renormalized compared to the underlying free band structure by means of spin-spin correlators, as we have quantified. Our main result is the establishment of a direct link between the dispersion of zeros and the transmission amplitude naturally probed in a co-tunneling experiment, which drastically simplifies the task of observing and characterizing Green's function zeros. \updt{Moreover, our approach allows, in principle, a reconstruction of the spectroscopically inaccessible \gls{gf} determinant from its individual matrix elements.} We have provided extensive numerical proof of the correctness and resilience of our approximation, and suggested a feasible experimental setup for the verification of our key findings. Numerical simulations suggest that the fingerprints of the zeros dispersion and renormalization are observable already for small systems: specifically, an ensemble of six quantum dots (2 leads and 4 in the central Hubbard tunneling medium) is already sufficient to quantify an interesting correlation induced renormalization of band parameters. 

\paragraph*{Acknowledgments}-- We acknowledge financial support from the German Research Foundation (DFG) through the Collaborative Research Centre SFB 1143 (Project-ID 247310070), the Cluster of Excellence ct.qmat (Project-ID 390858490). Our numerical calculations have been performed at the Center for Information Services and High Performance Computing (ZIH) at TU Dresden.

%%%%%%%%%%%%%%%%%%%%%%%%%%%%%%%%%%%%%%%%%%%%%%%%%%%%%%%%%%%%%%
%     BIBLIOGRAPHY
%%%%%%%%%%%%%%%%%%%%%%%%%%%%%%%%%%%%%%%%%%%%%%%%%%%%%%%%%%%%%%

%\phantomsection
\addcontentsline{toc}{chapter}{Bibliography}

\newpage
\onecolumngrid
\appendix
\section{Perturbation theory for the Green's function}
\label{SM:DDZ}
In the following section, we derive the \gls{gf} for the Hubbard model within the strongly correlated limit $\frac{D}{U}\rightarrow 0$ , i.e., deep in the Mott regime.
The general Hubbard Hamiltonian is given as:
\begin{align}
	H_{\text{H}} &= H_0 + U H_{\text{int}} = D \phi^{-1}\left( \phi H^{\prime}_0 + H_{\text{int}}\right) \label{EQ:sm_hubbard_ham} \\
	H_{\text{int}}  &= 2\sum_{j,\alpha} (n_{j,\downarrow} - \frac{1}{2})(n_{j,\uparrow} - \frac{1}{2})\text{,}
	\label{EQ:sm_hubbard_int}
\end{align}
where the non-interacting Hamiltonian $H_0 = \sum c^\dagger_{j,\sigma} h_{j,\sigma;j',\sigma'} c_{j',\sigma'}$ is quadratic in its creation(annihilation) operators $c^\dagger_{j,\sigma}$ ($c_{j,\sigma}$) and obeys the bandwidth $2D$. 
The Hubbard interaction $H_{\text{int}}$ consists on the local density operators $n_{j,\sigma}= c^\dagger_{j,\sigma}c_{j,\sigma} $, acting on the spinful lattice sites $j$.

Deep in the Mott regime, Eq.~\eqref{EQ:sm_hubbard_ham} implies an perturbative expansion \cite{Sakurai_2020} of the eigenstates in $\phi=\frac{D}{U}$, where the rescaled bare Hamiltonian $H^{\prime}_0 =  \frac{1}{D} H_0$ acts as perturbation. Then, we insert the resulting expanded eigenstates and eigenenergies into the Källen-Lehmann representation, apply an expansion at low energies and eventually collect all terms up to the second order in $\phi$.

\subsection{Eigenvector structure towards the strongly correlated limit}
\label{SM:evs}
\begin{figure}[htp]
	\centering
	\includegraphics[width=0.5\linewidth]{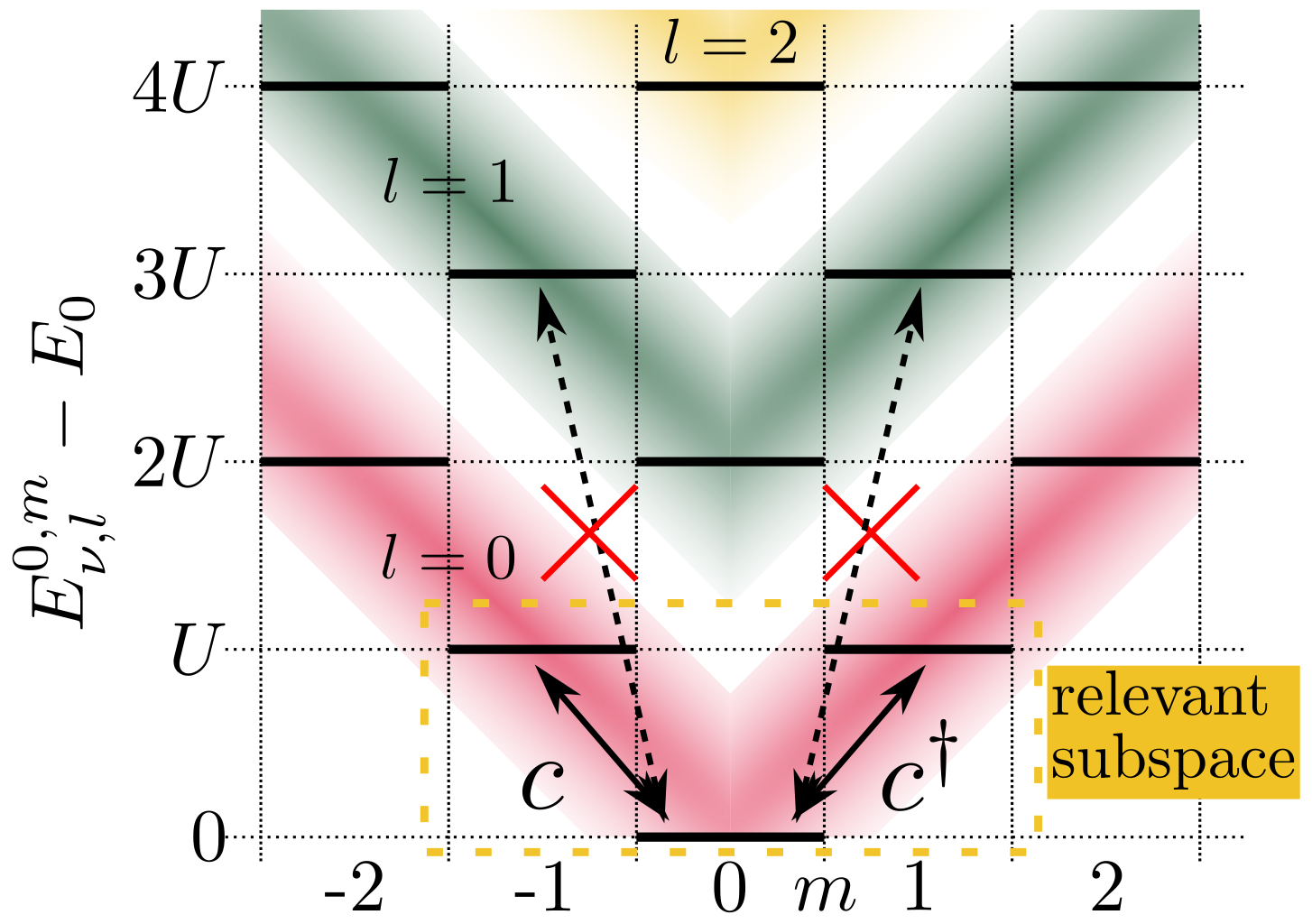}
	\caption{Energy levels $E^{0,m}_{\nu,l}= E_0 + U (\vert m \vert + 2l )$ for a typical Hubbard system in the Mott regime (atomic limit): all eigenstates can be sorted with respect to the particle-number deviation from half-filling $N=N_h + m$, and the amount of empty (douple occupied) sites $l$ additional to the lowest states for a fixed particle-number sector (coloured regions). For each energy there is a huge degeneracy denoted by $\nu$. The energy $E_0$ denotes the lowest global energy. In order to get the zero temperature \gls{gf}, the only relevant subspace consist of the global groundstate and the lowest energy sector with $m=\vert 1 \vert,l=0$ (yellow box). Direct Scattering from the groundstate into higher energies with $l>0$ is forbidden within the atomic limit (black arrows).}
	\label{fig:level_structure}
\end{figure}

The Hubbard interaction $H_{\text{int}}$ towards strong correlation $\phi \rightarrow 0$ avoids empty and double occupied sites, i.e., they cost an energy amount of $2D\phi^{-1}=2U$ . The eigenenergies will then arrange in well separated energy sectors according the amount of empty and double occupied sides (see Fig.\ref{fig:level_structure}).
We will expand the eigenstates around this strongly correlated limit, hence for house-holding reasons we will index-labelling all the unperturbed eigenstates $\{ \vert E^{0,m}_{\nu,l}\rangle\}$  according to their energy sectors.
In the following we denote by $m= N- N_h$ the relative particle number, consisting of the total particle number $N=\sum_{j,\sigma} n_{j,\sigma}$ and the particle number at half-filling $N_h$. 
Furthermore, by $l$ we count the empty or double occupied sites additional to the lowest energy state of the corresponding particle sector $m$.
In combination, the indices $l$ and $m$ fully determine the energy of each eigenstate in the atomic limit:
\begin{align}
	E^{0,m}_{\nu,l}= E_0 + U (\vert m \vert + 2l )\text{.}
\end{align} 
Here $E_0=-\frac{U}{2}N_s$ denotes a global offset, with the number of all spinful sites $N_s$.
The last index $\nu$ differentiates between the additional degenerated eigenstates within a particle sector $m,l$.

\subsection{Decomposition of the non-interacting Hamiltonian}
The bare hopping $H_0$ contains various terms, which can be resorted with respect to the energy-increase regarding the strongly interacting limit,i.e., their ability to produce empty and double-occupied sites by acting on the groundstate at half-filling $H_0 \vert GS_0 \rangle$.
\begin{align}
	H_0 = \sum_l H_{0,l}
\end{align}
Here $H_{0,l}$ scatters the groundstate $\vert GS_0 \rangle$ into the energy level $l$ within the half-filled particle sector $m=0$.

For the generic Hubbard model there are only local terms $H_{0,l=0}$($\equiv M$ in the letter), like on-site potentials and spin-flip terms, and non-local terms $H_{0,l=1}$($\equiv W$ in the letter), e.g., site-to-site hopping terms.

\subsection{Green's function approximation with SU(2)-symmetry}
\label{SM:GFA_SU2}
In this section we derive the low-energy \gls{gf} approximation for SU(2)-symmetric Mott insulators, the more general SU(2)-breaking case is treated in Section \ref{SM:GFA_woSU2}. We will first derive an rescaled \gls{gf} $ G^{\prime}(\phi\frac{\omega}{D}) = \frac{D}{\phi} G(\omega) $, where $G^{\prime}$ denotes the \gls{gf} corresponding to the Hamiltonian $D^{-1} \phi H_{\text{H}}$, afterwards we can easily readout $G(\omega)$.
We begin by formally expanding the perturbed eigenstates(-energies) $\vert E^{P,m}_{\nu,l} \rangle$($E^{P,m}_{\nu,l}$) of $\frac{\phi}{D} H_{\text{H}}$ in Eq.~\eqref{EQ:sm_hubbard_ham} towards the unperturbed limit, up to the first order in $\phi$. This is essentially standard degenerated perturbation theory~\cite{Sakurai_2020} and maps the atomic eigenstates(-energies) $\vert E^{0,m}_{\nu,l} \rangle$($E^{0,m}_{\nu,l}$) smoothly to the perturbed ones.
For the \gls{gf} approximation the following terms in the eigenstate expansion are sufficient \footnote{In general this expansion is incomplete, however, SU(2)-symmetry in combination with the first order expansion of the \gls{gf} suppresses lot's of terms.}:
\begin{align}
	E^{P,m}_{\nu,l} &\approx  E^{0,m}_{\nu,l} + \phi \langle  E^{0,m}_{\nu,l} \vert H^{\prime}_0 \vert E^{0,m}_{\nu,l} \rangle + \mathcal{O}(\phi^2)  \label{EQ:sm_expansion_e} \\
	\vert E^{P,m}_{\nu,l} \rangle 
	&\approx \vert E^{0,m}_{\nu,l} \rangle + \phi \sum_{l'\neq l}  \frac{P_{l'}^{m}  H^{\prime}_0 \vert E^{0,m}_{\nu,l} \rangle}{2(l - l')} + \mathcal{O}(\phi^2)  \label{EQ:sm_expansion} \\
	P_{l'}^{m} &=\sum_{\nu'}  \vert E^{0,m}_{\nu',l'} \rangle \langle  E^{0,m}_{\nu',l'} \vert  \nonumber
\end{align}
Here, we introduce the energy-sector projector $P_{l}^{m}$.
It is important to note that the degeneracy within each sector $(m,l)$, denoted by $\nu$, must, in principle, be resolved by applying non-trivial higher-order perturbation theory \cite{Sakurai_2020}. However, for the final result, it suffices that such an expansion exists. Moreover, Eq. \eqref{EQ:sm_expansion} can be interpreted as an inverse Schrieffer-Wolff transformation \cite{Schrieffer_1966, Bir_1974, Bravyi_2011}, which maps the eigenstates of an effective model, which is defined in the atomic limit, back to the original basis.

Let's insert the perturbed eigenstates(-energies) Eq.~\eqref{EQ:sm_expansion} into the Källen-Lehmann representation:  
\begin{align}
	G^{\prime}_{j,\sigma;j',\sigma'}(\phi \omega^{\prime}) &= \sum_{\nu,l} \Bigg[\frac{\langle GS^P |c_{j,\sigma}| E_{\nu,l}^{P,+} \rangle\langle  E_{\nu,l}^{P,+} |c^\dagger_{j',\sigma'}| GS^P \rangle}{\phi \omega^{\prime} + E_0^P - E_{\nu,l}^{P,+}} 
	+ \frac{\langle  E_{\nu,l}^{P,-} |c_{j,\sigma}| GS^P \rangle\langle GS^P |c^\dagger_{j',\sigma'}| E_{\nu,l}^{P,-} \rangle}{\phi \omega^{\prime} + E_{\nu,l}^{P,-} - E_0^P} \Bigg] \nonumber \\
	&= \sum_{\nu,l} \Bigg[\frac{A_{j,\sigma ; j',\sigma'}^{\nu,l}}{\phi \omega^{\prime} + E_0^P - E_{\nu,l}^{P,+}} 
	+ \frac{B_{j,\sigma ; j',\sigma'}^{\nu,l}}{\phi \omega^{\prime} + E_{\nu,l}^{P,-} - E_0^P} \Bigg]  \text{.} \label{SM:big_expansion}	
\end{align}

We already rescaled the frequency $\phi \omega^{\prime}$, where $\omega^{\prime}=\frac{\omega}{D}$.
We find for the expanded $A_{j,\sigma ; j',\sigma'}^{\nu,l}$:
\begin{align}
	A_{j,\sigma ; j',\sigma'}^{\nu,l} &=\langle  GS^P |c_{j,\sigma}| E^{P,+}_{\nu,l} \rangle  \langle  E^{P,+}_{\nu,l} |c^\dagger_{j',\sigma'}| GS^P \rangle \nonumber \\
	&\approx \langle  GS^0 \vert \left( 1 -  \sum_{k>0} \frac{\phi}{2k}   H^{\prime}_0 P^{0}_k \right)  c_{j,\sigma} \left( 1 + \sum_{l'\neq l} \frac{\phi}{2(l - l')} P^+_l H^{\prime}_0 \right)\vert E^{0,+}_{\nu,l} \rangle \nonumber \\ 
	&\times
	\langle   E^{0,+}_{\nu,l} \vert \left( 1 + \sum_{l'\neq l} \frac{\phi}{2(l - l')}  H^{\prime}_0 P^+_l \right)  c^\dagger_{j',\sigma'} \left( 1 -  \sum_{k>0} \frac{\phi}{2k}  P^{0}_k H^{\prime}_0  \right) \vert  GS^0 \rangle \nonumber \\
	&\approx \langle  GS^0 \vert c_{j,\sigma} + \phi \left[ -  \sum_{k>0} \frac{1}{2k}   H^{\prime}_0 P^{0}_k c_{j,\sigma}  +  c_{j,\sigma} \sum_{l'\neq l} \frac{1}{2(l - l')} P^+_{l'} H^{\prime}_0  \right]\vert E^{0,+}_{\nu,l} \rangle \nonumber \\ 
	&\times \langle  E^{0,+}_{\nu,l} \vert c^\dagger_{j',\sigma'} + \phi \left[ \sum_{l'\neq l} \frac{1}{2(l - l')}  H^{\prime}_0 P^+_{l'} c^\dagger_{j',\sigma'} - c^\dagger_{j',\sigma'} \sum_{k>0} \frac{1}{2k}  P^{0}_k H^{\prime}_0  \right] \vert  GS^0 \rangle \nonumber \\
	&\approx \langle  GS^0 \vert c_{j,\sigma} \vert E^{0,+}_{\nu,l} \rangle \langle  E^{0,+}_{\nu,l} \vert c^\dagger_{j',\sigma'}  \vert  GS^0 \rangle + \phi  \langle  GS^0 \vert c_{j,\sigma} \vert E^{0,+}_{\nu,l} \rangle \langle  E^{0,+}_{\nu,l} \vert  \sum_{l'\neq l} \frac{1}{2(l - l')}  H^{\prime}_0 P^+_{l'} c^\dagger_{j',\sigma'} \vert  GS^0 \rangle \nonumber \\
	&- \phi  \langle  GS^0 \vert c_{j,\sigma} \vert E^{0,+}_{\nu,l} \rangle \langle  E^{0,+}_{\nu,l} \vert c^\dagger_{j',\sigma'} \sum_{k>0} \frac{1}{2k}  P^{0}_k H^{\prime}_0   \vert  GS^0 \rangle - \phi \langle  GS^0 \vert   \sum_{k>0} \frac{1}{2k}   H^{\prime}_0 P^{0}_k c_{j,\sigma} \vert E^{0,+}_{\nu,l} \rangle \langle  E^{0,+}_{\nu,l} \vert c^\dagger_{j',\sigma'}  \vert  GS^0 \rangle \nonumber \\ 
	&+ \phi \langle  GS^0 \vert  c_{j,\sigma} \sum_{l'\neq l} \frac{1}{2(l - l')} P^+_{l'} H^{\prime}_0  \vert E^{0,+}_{\nu,l} \rangle \langle  E^{0,+}_{\nu,l} \vert c^\dagger_{j',\sigma'}  \vert  GS^0 \rangle  \nonumber \\
	&=A_{j,\sigma ; j',\sigma'}^{0,\nu,l=0} +\frac{\phi}{2} A_{j,\sigma ; j',\sigma'}^{1,\nu,l=0} + \mathcal{O}(\phi^2) \label{SM:exp_A}
\end{align}
Only the lowest energy subspace with $l=0$ is relevant for our expansion, since terms like $ \langle  GS^0 \vert c_{j,\sigma} \vert E^{0,+}_{\nu,l} \rangle, \langle  E^{0,+}_{\nu,l} \vert c^\dagger_{j,\sigma}  \vert  GS^0 \rangle$ are subspace conserving.
Similar we find for $ B_{j,\sigma ; j',\sigma'}^{\nu,l} $:
\begin{align}
	B_{j,\sigma ; j',\sigma'}^{\nu,l} &=\langle  GS^P |c^\dagger_{j',\sigma'} | E^{P,-}_{\nu,l} \rangle \langle  E^{P,-}_{\nu,l} | c_{j,\sigma}| GS^P \rangle \nonumber\\
	&\approx \langle  GS^0 \vert \left( 1 -  \sum_{k>0} \frac{\phi}{2k}   H^{\prime}_0 P^{0}_k \right)  c^\dagger_{j',\sigma'} \left( 1 + \sum_{l'\neq l} \frac{\phi}{2(l - l')} P^-_l H^{\prime}_0 \right)\vert E^{0,-}_{\nu,l} \rangle \nonumber \\ 
	&\times \langle  E^{0,-}_{\nu,l} \vert \left( 1 + \sum_{l'\neq l} \frac{\phi}{2(l - l')}  H^{\prime}_0 P^-_l \right)  c_{j,\sigma}  \left( 1 -  \sum_{k>0} \frac{\phi}{2k}  P^{0}_k H^{\prime}_0  \right) \vert  GS^0 \rangle \nonumber \\
	&\approx \langle  GS^0 \vert c^\dagger_{j',\sigma'} + \phi \left[ -  \sum_{k>0} \frac{1}{2k}   H^{\prime}_0 P^{0}_k  c^\dagger_{j',\sigma'} +  c^\dagger_{j',\sigma'} \sum_{l'\neq l} \frac{1}{2(l - l')} P^-_{l'} H^{\prime}_0  \right]\vert E^{0,-}_{\nu,l} \rangle \nonumber \\ 
	&\times \langle  E^{0,-}_{\nu,l} \vert c_{j,\sigma}   \phi \left[ \sum_{l'\neq l} \frac{1}{2(l - l')}  H^{\prime}_0 P^-_{l'} c_{j,\sigma}  - c_{j,\sigma} \sum_{k>0} \frac{1}{2k}  P^{0}_k H^{\prime}_0  \right] \vert  GS^0 \rangle \nonumber \\
	&\approx \langle  GS^0 \vert c^\dagger_{j',\sigma'} \vert E^{0,-}_{\nu,l} \rangle \langle  E^{0,-}_{\nu,l} \vert c_{j,\sigma}   \vert  GS^0 \rangle \nonumber \\ &+ \phi  \langle  GS^0 \vert c^\dagger_{j',\sigma'} \vert E^{0,-}_{\nu,l} \rangle \langle  E^{0,-}_{\nu,l} \vert  \sum_{l'\neq l} \frac{1}{2(l - l')}  H^{\prime}_0 P^-_{l'} c_{j,\sigma} \vert  GS^0 \rangle +  \phi \langle  GS^0 \vert c^\dagger_{j',\sigma'} \sum_{l'\neq l} \frac{1}{2(l - l')} P^-_{l'} H^{\prime}_0  \vert E^{0,-}_{\nu,l} \rangle \langle  E^{0,-}_{\nu,l} \vert c_{j,\sigma}  \vert  GS^0 \rangle \nonumber \\ 
	&-\phi  \langle  GS^0 \vert c^\dagger_{j',\sigma'} \vert E^{0,-}_{\nu,l} \rangle \langle  E^{0,-}_{\nu,l} \vert c_{j,\sigma} \sum_{k>0} \frac{1}{2k}  P^{0}_k H^{\prime}_0   \vert  GS^0 \rangle - \phi \langle  GS^0 \vert  \sum_{k>0} \frac{1}{2k}   H^{\prime}_0 P^{0}_k c^\dagger_{j',\sigma'} \vert E^{0,-}_{\nu,l} \rangle \langle  E^{0,-}_{\nu,l} \vert c_{j,\sigma}  \vert  GS^0 \rangle \nonumber \\
	&=B_{j,\sigma ; j',\sigma'}^{0,\nu,l=0} +\frac{\phi}{2} B_{j,\sigma ; j',\sigma'}^{1,\nu,l=0} + \mathcal{O}(\phi^2)  \label{SM:exp_B}
\end{align}

As next step, we expand the energies in the corresponding denominators with Eq.~\eqref{EQ:sm_expansion_e}:
\begin{align}
	\frac{1}{\phi \omega^{\prime} +E^P_0 - E^{P,+}_{\nu,l}}   
	\approx \frac{1}{\phi \omega^{\prime} -1 - 2l  + \phi \langle  GS^0 \vert H^{\prime}_0 \vert GS^0 \rangle   - \phi \langle  E^{0,+}_{\nu,l} \vert H^{\prime}_0 \vert E^{0,+}_{\nu,l} \rangle} \text{.}
\end{align}
Similarly, we get the second denominator:
\begin{align}
	\frac{1}{\phi \omega^{\prime} -E^P_0 + E^{P,-}_{\nu,l}} \approx \frac{1}{\phi \omega^{\prime} + 1 + 2l  - \phi \langle  GS^0 \vert H^{\prime}_0 \vert GS^0 \rangle   + \phi \langle  E^{0,-}_{\nu,l} \vert H^{\prime}_0 \vert E^{0,-}_{\nu,l} \rangle} \text{.}
\end{align}
Then, we expand the denominators at $\phi=0$. Notice that $l=0$ due to Eq.~\eqref{SM:exp_A} and Eq.~\eqref{SM:exp_B}:
\begin{align}
	\frac{A_{j,\sigma ; j',\sigma'}^{\nu,0}}{\phi \omega^{\prime} +E^P_0 - E^{P,+}_{\nu,0}}  \approx -A_{j,\sigma ; j',\sigma'}^{0,\nu,0} + \frac{\phi}{2} \left[  -A_{j,\sigma ; j',\sigma'}^{1,\nu,0}  - 2 A_{j,\sigma ; j',\sigma'}^{0,\nu,0} \left(\omega^{\prime}+ \langle  GS^0 \vert H^{\prime}_0 \vert GS^0 \rangle   -  \langle  E^{0,+}_{\nu,0} \vert H^{\prime}_0 \vert E^{0,+}_{\nu,0} \rangle\right) \right] + \mathcal{O}(\phi^2) \text{,} \label{SM:exp_A_tot}
\end{align} 
\begin{align}
	\frac{B_{j,\sigma ; j',\sigma'}^{\nu,0}}{\phi \omega^{\prime} -E^P_0 + E^{P,-}_{\nu,0}}  \approx B_{j,\sigma ; j',\sigma'}^{0,\nu,0} + \frac{\phi}{2}\left[ B_{j,\sigma ; j',\sigma'}^{1,\nu,0} + 2 B_{j,\sigma ; j',\sigma'}^{0,\nu,0} \left( -\omega^{\prime} + \langle  GS^0 \vert H^{\prime}_0 \vert GS^0 \rangle   - \langle  E^{0,-}_{\nu,0} \vert H^{\prime}_0 \vert E^{0,-}_{\nu,0} \rangle \right)\right] + \mathcal{O}(\phi^2) \text{.} \label{SM:exp_B_tot}
\end{align}
The last step involves expanding a rational function, which is valid only for frequencies $\vert \omega^{\prime} \vert \ll 1$, i.e., well within the Mott gap. Otherwise, performing this expansion near a pole would result in significant errors.
Including the summation in Eq.~\ref{SM:big_expansion} we can simplify each term even further:
\begin{align}
	-\sum_{\nu} A_{j,\sigma ; j',\sigma'}^{0,\nu,0} &= -\sum_{\nu} \langle  GS^0 \vert c_{j,\sigma} \vert E^{0,+}_{\nu,0} \rangle \langle  E^{0,+}_{\nu,0} \vert c_{j' ,\sigma'}^\dagger  \vert  GS^0 \rangle = -\langle  GS^0 \vert c_{j ,\sigma}  c_{j' ,\sigma'}^\dagger  \vert  GS^0 \rangle \label{SM:A00_0} \text{ ,} \\
	-\sum_{\nu} A_{j,\sigma ; j',\sigma'}^{1,\nu,0} &= \sum_{\nu}  \langle  GS^0 \vert c_{j ,\sigma} \vert E^{0,+}_{\nu,0} \rangle \langle  E^{0,+}_{\nu,0} \vert  \sum_{l'\neq 0} \frac{1}{l'}  H^{\prime}_0 P^+_{l'} c_{j' ,\sigma'}^\dagger + c_{j' ,\sigma'}^\dagger \sum_{k>0} \frac{1}{k}  P^{0}_k H^{\prime}_0   \vert  GS^0 \rangle \nonumber  \\
	&+ \langle  GS^0 \vert    \sum_{k>0} \frac{1}{k}   H^{\prime}_0 P^{0}_k c_{j ,\sigma} +  c_{j ,\sigma} \sum_{l'\neq 0} \frac{1}{ l'} P^+_{l'} H^{\prime}_0  \vert E^{0,+}_{\nu,0} \rangle \langle  E^{0,+}_{\nu,0} \vert c_{j' ,\sigma'}^\dagger  \vert  GS^0 \rangle \nonumber \\
	&=   \langle  GS^0 \vert c_{j ,\sigma} c_{j' ,\sigma'}^\dagger  \sum_{k>0} \frac{1}{k}P^{0}_k  H^{\prime}_0   \vert  GS^0 \rangle + \langle  GS^0 \vert     H^{\prime}_0 \sum_{k>0} \frac{1}{k}P^{0}_k  c_{j ,\sigma} c_{j' ,\sigma'}^\dagger  \vert  GS^0 \rangle \nonumber \\
	&=   \sum_{k>0} \frac{1}{k}\langle  GS^0 \vert \left[ c_{j ,\sigma} c_{j' ,\sigma'}^\dagger,    H^{\prime}_{0,k} \right]_+ \vert  GS^0 \rangle \text{ ,} \label{SM:A10} \\	
	\sum_{\nu}  A_{j,\sigma ; j',\sigma'}^{0,\nu,0} \langle  E^{0,+}_{\nu,0} \vert H^{\prime}_0 \vert E^{0,+}_{\nu,0} \rangle   
	&= \sum_{\nu}  \langle  GS^0 \vert c_{j ,\sigma} \vert E^{0,+}_{\nu,0} \rangle \langle  E^{0,+}_{\nu,0} \vert c_{j' ,\sigma'}^\dagger  \vert  GS^0 \rangle  \langle  E^{0,+}_{\nu,0} \vert H^{\prime}_0 \vert E^{0,+}_{\nu,0} \rangle  \nonumber \\
	&=   \langle  GS^0 \vert c_{j ,\sigma} H^{\prime}_0 c_{j' ,\sigma'}^\dagger  \vert  GS^0 \rangle \text{.} \label{SM:A00_1}
\end{align}
We find similar results for the $B$-terms:
\begin{align}
	\sum_{\nu} B_{j,\sigma ; j',\sigma'}^{0,\nu,0} &= \sum_{\nu} \langle  GS^0 \vert c_{j' ,\sigma'}^\dagger \vert E^{0,-}_{\nu,0} \rangle \langle  E^{0,-}_{\nu,0} \vert   c_{j ,\sigma}\vert  GS^0 \rangle = \langle  GS^0 \vert c_{j' ,\sigma'}^\dagger   c_{j ,\sigma}\vert  GS^0 \rangle \text{ ,} \label{SM:B00_0} \\
	\sum_{\nu} B_{j,\sigma ; j',\sigma'}^{1,\nu,0}  &= -\sum_{\nu}  \langle  GS^0 \vert c_{j' ,\sigma'}^\dagger \vert E^{0,-}_{\nu,0} \rangle \langle  E^{0,-}_{\nu,0} \vert  \sum_{l'\neq 0} \frac{1}{ l'}  H^{\prime}_0 P^-_{l'} c_{j ,\sigma} + c_{j ,\sigma} \sum_{k>0} \frac{1}{k}  P^{0}_k H^{\prime}_0   \vert  GS^0 \rangle \nonumber \\
	&- \sum_{\nu} \langle  GS^0 \vert    \sum_{k>0} \frac{1}{k}   H^{\prime}_0 P^{0}_k c_{j' ,\sigma'}^\dagger +  c_{j' ,\sigma'}^\dagger \sum_{l'\neq 0} \frac{1}{ l'} P^-_{l'} H^{\prime}_0  \vert E^{0,-}_{\nu,0} \rangle \langle  E^{0,-}_{\nu,0} \vert c_{j ,\sigma}   \vert  GS^0 \rangle \nonumber \\
	&=   -\sum_{k>0}\frac{1}{k} \langle  GS^0 \vert \left[ c_{j' ,\sigma'}^\dagger c_{j ,\sigma},  H^{\prime}_{0,k}   \right]_+ \vert  GS^0 \rangle \text{ ,} \label{SM:B10} \\
	\sum_{\nu}  B_{j,\sigma ; j',\sigma'}^{0,\nu,0}  \langle  E^{0,-}_{\nu,0} \vert H^{\prime}_0 \vert E^{0,-}_{\nu,0} \rangle  
	&= \sum_{\nu}  \langle  GS^0 \vert c_{j' ,\sigma'}^\dagger \vert E^{0,-}_{\nu,0} \rangle \langle  E^{0,-}_{\nu,0} \vert  c_{j ,\sigma} \vert  GS^0 \rangle  \langle  E^{0,-}_{\nu,0} \vert H^{\prime}_0 \vert E^{0,-}_{\nu,0} \rangle  \nonumber \\
	&=   \langle  GS^0 \vert c_{j' ,\sigma'}^\dagger H^{\prime}_0  c_{j ,\sigma} \vert  GS^0 \rangle   \text{.}  \label{SM:B00_1}	
\end{align}
All terms together: Eq.~\eqref{SM:big_expansion}, Eq.~\eqref{SM:exp_A_tot},Eq.~\eqref{SM:exp_B_tot}, Eq.~\eqref{SM:A00_0}, Eq.~\eqref{SM:A10}, Eq.~\eqref{SM:A00_1}, Eq.~\eqref{SM:B00_0}, Eq.~\eqref{SM:B10}, Eq.~\eqref{SM:B00_1}, result in the first order approximation of the \gls{gf}:
\begin{align}
	G^{\prime}_{j,\sigma ; j',\sigma'}(\phi \omega^{\prime}) &\approx \left(G^{\prime}_0\right)_{j,\sigma ; j',\sigma'} + \phi \left(G^{\prime}_1\right)_{j,\sigma ; j',\sigma'} - \phi \omega^{\prime} + \mathcal{O}(\phi^2)	  \\
	\left(G^{\prime}_0\right)_{j,\sigma ; j',\sigma'} &= \langle  GS^0 \vert c_{j' ,\sigma'}^\dagger c_{j ,\sigma}  \vert  GS^0 \rangle  -\langle  GS^0 \vert c_{j ,\sigma}c_{j' ,\sigma'}^\dagger  \vert  GS^0 \rangle  \label{EQ:sm_tg0}\\
	\left(G^{\prime}_1\right)_{j,\sigma ; j',\sigma'} &= \sum_{k>0}\frac{1}{2k} \left[  \langle  GS^0 \vert \left[ c_{j ,\sigma} c_{j' ,\sigma'}^\dagger,    H^{\prime}_{0,k} \right]_+ \vert  GS^0 \rangle  - \langle  GS^0 \vert \left[ c_{j' ,\sigma'}^\dagger c_{j ,\sigma},  H^{\prime}_{0,k}   \right]_+ \vert  GS^0 \rangle\right] \nonumber \\ 
	&\quad + \left[  \langle  GS^0 \vert c_{j ,\sigma}(H^{\prime}_0 -\Delta^{\prime}_0)c_{j' ,\sigma'}^\dagger  \vert  GS^0 \rangle  -  \langle  GS^0 \vert c_{j' ,\sigma'}^\dagger (H^{\prime}_0-\Delta^{\prime}_0) c_{j ,\sigma}  \vert  GS^0 \rangle \right] \label{EQ:sm_tg1} \\
	\Delta^{\prime}_0&=\langle  GS^0 \vert H^{\prime}_0 \vert  GS^0 \rangle
	\label{SM:exp_gf1}
\end{align}
Since $k>0$ and due to the single occupation constrain we can simplify:
\begin{align}
	\langle  GS^0 \vert \left[ c_{j' ,\sigma'}^\dagger c_{j ,\sigma},  H^{\prime}_{0,k}   \right]_+ \vert  GS^0 \rangle = -\langle  GS^0 \vert \left[ c_{j ,\sigma} c_{j' ,\sigma'}^\dagger,    H^{\prime}_{0,k} \right]_+ \vert  GS^0 \rangle \text{.}
\end{align}
Hence we find:
\begin{align}
	\left(G^{\prime}_0\right)_{j,\sigma ; j',\sigma'} &= \sum_{k>0}\frac{1}{k}  \langle  GS^0 \vert \left[ c_{j ,\sigma} c_{j' ,\sigma'}^\dagger,    H^{\prime}_{0,k} \right]_+ \vert  GS^0 \rangle \nonumber \\ 
	&\quad + \left[ \langle  GS^0 \vert c_{j ,\sigma}(H^{\prime}_0-\Delta^{\prime}_0)c_{j' ,\sigma'}^\dagger  \vert  GS^0 \rangle  -  \langle  GS^0 \vert c_{j' ,\sigma'}^\dagger (H^{\prime}_0-\Delta^{\prime}_0) c_{j ,\sigma}  \vert  GS^0 \rangle \right]  \text{.}
\end{align}
Notice that for the generic Hubbard model $k\leq 1$, i.e., the non-interacting Hamiltonian $H^{\prime}_0$ can be fully decomposed into local terms $H^{\prime}_{0,l=0}$ and non-local terms $H^{\prime}_{0,l=1}$.
As final step we do the rescaling, $G(\omega) = \frac{\phi}{D} G^{\prime}(\phi\frac{\omega}{D})$, and introduce the Hamiltonian of zeros $H^{\text{z}}$:
\begin{align}
	G(\omega) &\approx \frac{\phi^2}{D^2}  \left[H^{\text{z}} - \omega\right] + \mathcal{O}(\phi^3) \text{ ,} \\
	H^{\text{z}} &=  D\phi^{-1} G^0 +  G^1
	\text{.} \label{EQ:sm_gf_hubbard}
\end{align}
where  $G^0=G^{\prime}_0$ and $G^1=DG^{\prime}_1$.
Surprisingly, the only remaining non-trivial task is to evaluate the correct groundstate $\vert GS^0 \rangle $ towards the atomic limit $\phi \rightarrow 0 $ (that is degenerated perturbation theory~\cite{Sakurai_2020}), which is at half-filling generically the groundstate of the corresponding low-energy effective spin-$1\slash2$ Hamiltonian.

\subsection{Rewriting the GF-approximation using spin operators}
\label{SM:rWGFA_SU2}
The non-trivial groundstate $\vert GS^0 \rangle $ at half-filling lives in the subspace of singly occupied sites ($m=0,l=0$), what allows to rewrite the creation(annihilation)-operator strings in Eq.~\eqref{EQ:sm_tg0} and Eq.~\eqref{EQ:sm_tg1} as spin operators (Jordan-Wigner transformation).
The spin representation is defined as:
\begin{align}
	\overrightarrow{S}_j= \frac{1}{2}\Bigg(c^\dagger_{j,\uparrow} c_{j,\downarrow}+ c^\dagger_{j,\downarrow} c_{j,\uparrow},-i(c^\dagger_{j,\uparrow} c_{j,\downarrow}- c^\dagger_{j,\downarrow} c_{j,\uparrow}),
	n_{j,\uparrow} - n_{j,\downarrow}\Bigg)\text{.} \label{EQ:spin_rep}
\end{align}
Furthermore, within the singly occupied subspace we introduce the following useful identities, $\overrightarrow{\tau}=(\tau_x,\tau_y,\tau_z)$ denotes the Pauli matrices:
\begin{align}
	c_{j,\sigma} c^\dagger_{j,\sigma'} = \left( c_{j} c^\dagger_{j}  \right)_{\sigma,\sigma'} = \left( \frac{1}{2} \tau_0 - \overrightarrow{S}_j \cdot \overrightarrow{\tau} \right)_{\sigma,\sigma'} \text{,} \\
	c^\dagger_{j,\sigma'} c_{j,\sigma} = \left( c^\dagger_{j} c_{j}  \right)_{\sigma,\sigma'} = \left(\frac{1}{2} \tau_0 + \overrightarrow{S}_j \cdot \overrightarrow{\tau} \right)_{\sigma,\sigma'} \text{.}
\end{align}
The zeroth order $G^0$ is fully independent of the non-interacting Hamiltonian $H_0$ and we can directly apply the previous identities:
\begin{align}
	G^0_{j,\sigma;j',\sigma'} &= 2\langle   \overrightarrow{S}_j \rangle \cdot \overrightarrow{\tau} \delta_{j,j'}  \text{.}
\end{align}

For the first order contribution $G^1$, the  non-interacting Hamiltonian $H_0$ plays a crucial role. In the following, we provide a detailed derivation for a common bare Hamiltonian $H_0$. We begin by decomposing $G^1$ in Eq.~\eqref{EQ:sm_tg1} as:
\begin{align}
	G^1 &= L + R - W  \label{EQ:sm_g1_construction}\\
	L_{j,\sigma ; j',\sigma'} &= \sum_{k>0}\frac{1}{2k} \left[  \langle  GS^0 \vert \left[ c_{j ,\sigma} c_{j' ,\sigma'}^\dagger,    H_{0,k} \right]_+ \vert  GS^0 \rangle  - \langle  GS^0 \vert \left[ c_{j' ,\sigma'}^\dagger c_{j ,\sigma},  H_{0,k}   \right]_+ \vert  GS^0 \rangle\right] \label{EQ:sm_L} \\
	R_{j,\sigma ; j',\sigma'} &=   \langle  GS^0 \vert c_{j ,\sigma}(H_0 -\Delta_0)c_{j' ,\sigma'}^\dagger  \vert  GS^0 \rangle \label{EQ:sm_R} \\
	V_{j,\sigma ; j',\sigma'} &= \langle  GS^0 \vert c_{j' ,\sigma'}^\dagger (H_0-\Delta_0) c_{j ,\sigma}  \vert  GS^0 \rangle \label{EQ:sm_W}
\end{align}
Each term in $H_0$ can be studied separately, due to the linearity one can add all corresponding $G^1$ in the final result.
First we investigate terms acting as arbitrary on-site potentials $M$ ($\equiv H_{0,l=0}$ ):
\begin{align}
	M &= \sum_{k}\bar{\mu}_{k}  (n_{k,\uparrow} + n_{k,\downarrow}) \text{.} \label{EQ:sm_def_m}
\end{align}
Here, we obtain the local spin-isotropic potentials $\bar{\mu}_{k} $.
The individual components of $G^1$ (compare to Eqs.(\ref{EQ:sm_L}-\ref{EQ:sm_W})) can be simplified as: 
\begin{align}
	\Delta_0 &= \langle  M \rangle = \sum_{k}\bar{\mu}_{k} \\ 
	L_{j,\sigma ; j',\sigma'} &= 0 \\
	R_{j,\sigma ; j',\sigma'} &= \langle c_{j ,\sigma}(M-\Delta_0)c_{j' ,\sigma'}^\dagger  \rangle = \sum_k \bar{\mu}_{k} \langle c_{j ,\sigma}\left[\left(n_{k,\uparrow} + n_{k,\downarrow}\right)-1\right]c_{j' ,\sigma'}^\dagger  \rangle  \nonumber \\
	&= \sum_k \bar{\mu}_{k} \langle c_{j ,\sigma}\left[n_{k,\uparrow} + n_{k,\downarrow}, c_{j' ,\sigma'}^\dagger \right] \rangle = \bar{\mu}_{j} \delta_{j , j'} \left( \frac{1}{2} \tau_0 - \overrightarrow{\tau} \cdot \langle \overrightarrow{S}_{j}\rangle\right)_{\sigma,\sigma'} 
	\label{EQ:sm_rewr_ons_a} \\
	V_{j,\sigma ; j',\sigma'} &= \langle c_{j' ,\sigma'}^\dagger (M-\Delta_0) c_{j ,\sigma} \rangle  = \sum_k \bar{\mu}_{k} \langle c_{j' ,\sigma'}^\dagger \left[  \left(n_{k,\uparrow} + n_{k,\downarrow}\right)-1\right] c_{j ,\sigma} \rangle  \nonumber \\
	&= \sum_k \bar{\mu}_{k} \langle c_{j' ,\sigma'}^\dagger \left[n_{k,\uparrow} + n_{k,\downarrow},  c_{j ,\sigma} \right] \rangle  = -\bar{\mu}_{j} \delta_{j , j'} \left( \frac{1}{2} \tau_0 + \overrightarrow{\tau}  \cdot \langle \overrightarrow{S}_{j}\rangle\right)_{\sigma,\sigma'} 
\end{align}
Finally, we find for $G^1$ (add all terms as in Eq.~\eqref{EQ:sm_g1_construction}):
\begin{align}
	G^1_{j,\sigma ; j',\sigma'} &=  \delta_{j , j'}  \bar{\mu}_{j'} \left( \tau_0 \right)_{\sigma,\sigma'}  =  M_{j,\sigma ; j',\sigma'}  
	\label{EQ:sm_rewr_ons_b}
\end{align}
We find that for $SU(2)$-symmetric on-site potentials, $G^1$ equals $M$.
As second bare hopping Hamiltonian we investigate a typical spin-isotropic hopping $W$($\equiv H_{0,l=1}$):
\begin{align}
	W = \frac{1}{2} \sum_{k,k',\rho} t_{k,k'} c_{k,\rho}^\dagger c_{k',\rho} + \text{h.c. ,} \quad  t_{k,k} = 0   
\end{align}

\begin{align}
	L_{j,\sigma;j',\sigma'} &= \frac{1}{2} \sum_{k,k',\rho} \left( t_{k,k'} \langle  \left[ c_{j,\sigma} c^\dagger_{j',\sigma'}, c_{k,\rho}^\dagger c_{k',\rho} \right]_+ \rangle  +  t^\ast_{k,k'} \langle  \left[ c_{j,\sigma} c^\dagger_{j',\sigma'}, c_{k',\rho}^\dagger c_{k,\rho} \right]_+ \rangle \right)  \nonumber \\
	&= \sum_{\rho}  t_{j , j'} \langle  \left[ c_{j,\sigma} c^\dagger_{j',\sigma'}, c_{j,\rho}^\dagger c_{j',\rho} \right]_+ \rangle  \nonumber \\
	&= -\sum_{\rho}  t_{j , j'} \left[ \langle   c_{j\sigma}  c_{j,\rho}^\dagger c^\dagger_{j',\sigma'} c_{j',\rho}  \rangle  +  \langle  c_{j,\rho}^\dagger  c_{j,\sigma} c_{j',\rho}  c^\dagger_{j',\sigma'}   \rangle \right] \nonumber \\
	&= -\sum_{\rho}  t_{j , j'} \left[ \langle   \left( \frac{1}{2} \tau_0 - \overrightarrow{S}_{j} \cdot \overrightarrow{\tau} \right)_{\sigma\rho} \left( \frac{1}{2} \tau_0 + \overrightarrow{S}_{j'} \cdot \overrightarrow{\tau} \right)_{\rho \sigma'}  \rangle  +  \langle  \left( \frac{1}{2} \tau_0 + \overrightarrow{S}_{j} \cdot \overrightarrow{\tau} \right)_{\sigma\rho}  \left( \frac{1}{2} \tau_0 - \overrightarrow{S}_{j'} \cdot \overrightarrow{\tau} \right)_{\rho\sigma'}   \rangle\right]  \nonumber \\
	&= -  t_{j , j'} \left[ \langle  \left(\frac{1}{2}  - 2 \overrightarrow{S}_{j} \cdot \overrightarrow{S}_{j'} \right)\tau_0  -2i \left( \overrightarrow{S}_{j} \times \overrightarrow{S}_{j'}  \right) \cdot \overrightarrow{\tau}  \rangle\right]_{\sigma \sigma'}
	\label{EQ:sm_rewr_nn_t}
\end{align}

\begin{align}
	R_{j,\sigma;j',\sigma'} &= \frac{1}{2} \sum_{k,k',\rho} \left( t_{k,k'} \langle   c_{j,\sigma} c_{k,\rho}^\dagger c_{k',\rho} c^\dagger_{j',\sigma'} \rangle  +  t^\ast_{k,k'} \langle   c_{j,\sigma}  c_{k',\rho}^\dagger c_{k,\rho} c^\dagger_{j',\sigma'}  \rangle \right)  \nonumber \\
	&=  \sum_{\rho} t_{j , j'} \langle   \left( \frac{1}{2} \tau_0 - \overrightarrow{S}_{j} \cdot\overrightarrow{\tau} \right)_{\sigma\rho} \left( \frac{1}{2} \tau_0 - \overrightarrow{S}_{j'} \cdot \overrightarrow{\tau} \right)_{\rho\sigma'} \rangle  \nonumber \\
	&=  t_{j , j'} \left[ \langle   \left( \frac{1}{4}   + \overrightarrow{S}_{j} \cdot \overrightarrow{S}_{j'} \right) \tau_0 + \left( i \left( \overrightarrow{S}_{j} \times \overrightarrow{S}_{j'} \right) \cdot \overrightarrow{\tau} - \frac{1}{2}\overrightarrow{S}_{j'} \cdot \overrightarrow{\tau} - \frac{1}{2}\overrightarrow{S}_{j} \cdot \overrightarrow{\tau} \right)  \rangle \right]_{\sigma \sigma'} 
	\label{EQ:sm_rewr_nn_r}
\end{align}

\begin{align}
	V_{j,\sigma;j',\sigma'} &= \frac{1}{2} \sum_{k,k',\rho} \left( t_{k,k'} \langle    c^\dagger_{j',\sigma'} c_{k,\rho}^\dagger c_{k',\rho} c_{j,\sigma} \rangle  +  t^\ast_{k,k'} \langle    c^\dagger_{j',\sigma'}  c_{k',\rho}^\dagger c_{k,\rho} c_{j,\sigma} \rangle \right)  \nonumber \\
	&=  - \sum_{\rho} t_{j , j'} \langle    c_{j,\rho}^\dagger  c_{j,\sigma} c^\dagger_{j',\sigma'}  c_{j',\rho} \rangle  \nonumber \\
	&=  -\sum_{\rho} t_{j , j'} \langle   \left( \frac{1}{2} \tau_0 + \overrightarrow{S}_{j} \cdot \overrightarrow{\tau} \right)_{\sigma\rho} \left( \frac{1}{2} \tau_0 + \overrightarrow{S}_{j'} \cdot \overrightarrow{\tau} \right)_{\rho\sigma'} \rangle  \nonumber \\
	&=  -t_{j , j'} \left[ \langle   \left( \frac{1}{4}   + \overrightarrow{S}_{j} \cdot \overrightarrow{S}_{j'} \right) \tau_0 + \left( i \left( \overrightarrow{S}_{j} \times \overrightarrow{S}_{j'} \right) \cdot \overrightarrow{\tau} + \frac{1}{2}\overrightarrow{S}_{j'} \cdot \overrightarrow{\tau} + \frac{1}{2}\overrightarrow{S}_{j} \cdot \overrightarrow{\tau} \right)  \rangle \right]_{\sigma \sigma'}
	\label{EQ:sm_rewr_nn_w}
\end{align}
In total we find:
\begin{align}
	G^1_{j,\sigma ; j',\sigma'} &= 4 t_{j , j'} \left[ \langle  \overrightarrow{S}_{j} \cdot \overrightarrow{S}_{j'} \rangle \tau_0+ i \langle  \overrightarrow{S}_{j} \times \overrightarrow{S}_{j'} \rangle \cdot \overrightarrow{\tau} \right]_{\sigma,\sigma'}  
\end{align}

The final expression for $G^1$ corresponding to the total bare Hamiltonian, $H_0 = W +  M$, is obtained by summing all individual contributions to $G^1$. Using $G^0$ and $G^1$, we can determine the Hamiltonian of zeros, $H^{\text{z}}$.
For a spin-isotropic $H_0$ we find:
\begin{align}
	H^{\text{z}}_{j,\sigma ; j',\sigma'} = \left[ \left( 4  W_{j , j'}\langle  \overrightarrow{S}_{j} \cdot \overrightarrow{S}_{j'} \rangle_{j\neq j'}  + M_{j , j'} \delta_{j , j'} \right)\tau_0    + 4iW_{j , j'} \langle  \overrightarrow{S}_{j} \times \overrightarrow{S}_{j'} \rangle_{j\neq j'} \cdot \overrightarrow{\tau}   + 2 D\phi^{-1} \delta_{j , j'}\langle  \overrightarrow{S}_j  \rangle \cdot \overrightarrow{\tau} \right]_{\sigma,\sigma'}
\end{align}
For brevity, we omit the spin indices of the spin-isotropic bare Hamiltonian, where $(H_0)_{j,\uparrow;j',\uparrow}= (H_0)_{j,\downarrow;j',\downarrow} =(H_0)_{j,j'}$.
Furthermore, by construction, $M$ is strictly local, satisfying $M_{j;j'} = M_{j;j} \delta_{j,j'}$, while $W$ is  strictly non-local, $W_{j;j}=0$.
For an $SU(2)$-symmetric Hamiltonian of finite size, the groundstate typically retains $SU(2)$-symmetry. In this case, the local spin polarization vanishes, $\langle  \overrightarrow{S}_j  \rangle = 0$. 
Additionally, the spin cross-product also vanishes, $\langle  \overrightarrow{S}_{j} \times \overrightarrow{S}_{j'} \rangle = 0$, as this follows from its invariance under global spin rotations. Thus, for an $SU(2)$-symmetric Hamiltonian with $SU(2)$-symmetric groundstate, we obtain the elegant result presented in our letter:
\begin{align}
	H^{\text{z}}_{j,\sigma ; j',\sigma'} = \left( 4  W_{j , j'}\langle  \overrightarrow{S}_{j} \cdot \overrightarrow{S}_{j'} \rangle_{j\neq j'}  + M_{j , j'}  \delta_{j,j'} \right) \delta_{\sigma,\sigma'}
	\label{EQ:sm_final_zh}
\end{align}
As a final result, and in line with~\cite{Wagner_2023}, we observe that the local terms of the bare Hamiltonian $M$ reemerge unrenormalized in the $H^z$-Hamiltonian, while the non-local terms $W$ are rescaled by the spin correlation function at the corresponding bond, $\langle  \overrightarrow{S}_{j} \cdot \overrightarrow{S}_{j'} \rangle_{j\neq j'}$. The expectation values are taken with respect to the groundstate $\ket{GS_0}$ in the unperturbed limit, i.e.,  the groundstate of the corresponding effective spin-$1\slash2$ model (degenerated perturbation theory). Although $\ket{GS_0}$ is defined within the subspace of the atomic limit, it remains generally non-trivial, i.e., long-range correlations can occur:
\begin{align}
	\lim_{\phi\rightarrow 0} \langle  \overrightarrow{S}_j \cdot \overrightarrow{S}_{j'} \rangle \neq \langle  \overrightarrow{S}_j \cdot \overrightarrow{S}_{j'} \rangle_{\phi=0} \text{.}
\end{align}
Due to the additive structure of the \gls{gf}, Eq.~\eqref{SM:big_expansion}, our results are valid for any density matrix $\rho=\sum_\nu w_\nu \vert E^{P,m}_{\nu,l} \rangle \langle   E^{P,m}_{\nu,l} \vert $, which contains only SU(2)-symmetric states from the lowest energy sector $m = 0,~l=0$.  
In that case, all the expectation values can be simply updated:
\begin{align}
	\langle  ~ \cdot ~ \rangle = \langle  GS^0 \vert  \cdot \vert GS^0  \rangle \rightarrow \text{Tr}\left[ \rho ~ \cdot ~ \right] \text{.}
\end{align}

\subsection{Green's function approximation without SU(2)-symmetry}
\label{SM:GFA_woSU2}
In the the previous section we explicitly assumed SU(2)-symmetry, what allows to truncate the perturbative series of the eigenstates and gain a rather simple expression on the \gls{gfz}-Hamiltonian. However, in order to derive the Green's function approximation SU(2)-symmetry is not necessary.
The Källen-Lehmann presentation of the Green's function can be given as:
\begin{align}
	G_{j,\sigma;j',\sigma'}( \omega) &= \sum_{\nu} \Bigg[\frac{\langle GS^P |c_{j,\sigma}| E_{\nu}^{P} \rangle\langle E_{\nu}^{P} |c^\dagger_{j',\sigma'}| GS^P \rangle}{ \omega + E_0^P - E_{\nu}^{P}} 
	+ \frac{\langle E_{\nu}^{P} |c_{j,\sigma}| GS^P \rangle\langle GS^P |c^\dagger_{j',\sigma'}| E_{\nu}^{P} \rangle}{ \omega + E_{\nu}^{P} - E_0^P} \Bigg]  \\
	&= \mathcal{A}_{j,\sigma;j',\sigma'} + \mathcal{B}_{j,\sigma;j',\sigma'}\label{SM:big_expansion_2}
\end{align}
for brevity we supress all the sub-indices, and sum over all eigenstates $\{\vert E_\nu \rangle\}$. In the following we will treat the GF perturbatively in $\phi$ and in the frequency $\omega$, both up to its first order. We send $\phi \rightarrow 0$, mathematically correct we thereby assume a fixed $U$ and $D \rightarrow 0$, this is however equivalent to the strongly correlated limit. Due to the similarity, we only give details on the expansion for the first term in Eq.~\eqref{SM:big_expansion_2}.
As first step, we apply the expansion in $\phi$:
\begin{align}
	\mathcal{A}(\phi) \approx \mathcal{A}\big|_{\phi=0} +  \phi \partial_{\phi}\mathcal{A}\big|_{\phi=0}  
\end{align}
The individual terms are:
\begin{align}
	\mathcal{A}\big|_{\phi=0} &= \sum_{\nu} \frac{\langle GS^0 |c_{j,\sigma}| E_{\nu}^{0} \rangle\langle E_{\nu}^{0} |c^\dagger_{j',\sigma'}| GS^0 \rangle}{ \omega + E_0^0 - E_{\nu}^{0}} \\
	\partial_{\phi}\mathcal{A}\big|_{\phi=0}  &=  \sum_{\nu} \bigg[\frac{\langle GS^0 |c_{j,\sigma}\partial_{\phi}\left[| E_{\nu}^{P} \rangle\langle E_{\nu}^{P} |\right]\big|_{\phi=0}c^\dagger_{j',\sigma'}| GS^0 \rangle}{ \omega + E_0^0 - E_{\nu}^{0}} +  \frac{\langle E_{\nu}^{0} |c^\dagger_{j',\sigma'}\partial_{\phi} \left[| GS^P \rangle\langle GS^P |\right]\big|_{\phi=0} c_{j,\sigma}| E_{\nu}^{0} \rangle}{ \omega + E_0^0 - E_{\nu}^{0}} \\ 
	&-  \frac{\langle GS^0 |c_{j,\sigma}| E_{\nu}^{0} \rangle\langle E_{\nu}^0 |c^\dagger_{j',\sigma'}| GS^0 \rangle}{( \omega + E_0^0 - E_{\nu}^{0})^2}  \partial_{\phi} \left[E_0^P - E_{\nu}^{P}\right]\big|_{\phi=0}\bigg]
\end{align}
In each term we recognize amplitudes like $\langle E_{\nu}^0 |c^{(\dagger)}_{j',\sigma'}| GS^0 \rangle$ which in combination with the unperturbed groundstate $\vert GS^0 \rangle $ is non-vanishing only in the lowest energy sector (see Section \ref{SM:evs}). Most important hereby is that the denominator becomes constant:
\begin{align}
	\mathcal{A}\big|_{\phi=0} &= \sum_{\nu} \frac{\langle GS^0 |c_{j,\sigma}| E_{\nu}^{0} \rangle\langle E_{\nu}^{0} |c^\dagger_{j',\sigma'}| GS^0 \rangle}{ \omega - U} \label{SM:A_zo}\\
	\partial_{\phi}\mathcal{A}\big|_{\phi=0}  &=  \sum_{\nu} \frac{\langle GS^0 |c_{j,\sigma}\partial_{\phi}\left[| E_{\nu}^{P} \rangle\langle E_{\nu}^{P} |\right]\big|_{\phi=0}c^\dagger_{j',\sigma'}| GS^0 \rangle}{ \omega -U} +  \frac{\langle E_{\nu}^{0} |c^\dagger_{j',\sigma'}\partial_{\phi} \left[| GS^P \rangle\langle GS^P |\right]\big|_{\phi=0} c_{j,\sigma}| E_{\nu}^{0} \rangle}{ \omega -U} \\ 
	&-  \frac{\langle GS^0 |c_{j,\sigma}| E_{\nu}^{0} \rangle\langle E_{\nu}^0 |c^\dagger_{j',\sigma'}| GS^0 \rangle}{( \omega -U)^2}  \partial_{\phi} \left[E_0^P - E_{\nu}^{P}\right]\Big|_{\phi=0} \label{SM:A_fo}
\end{align}
In the next step we want to take the sum over all eigenstates $\{\vert E_\nu \rangle\}$. Here, Eq.~\eqref{SM:A_zo} can be easily evaluated:
\begin{align}
	\mathcal{A}\big|_{\phi=0} &=  \frac{\langle GS^0 |c_{j,\sigma}c^\dagger_{j',\sigma'}| GS^0 \rangle}{ \omega - U}
\end{align}
In Eq.~\eqref{SM:A_fo}, the first term will vanish, what can be seen by using that the summation and the derivative commutes:
\begin{align}
	\sum_{\nu} \partial_{\phi}\left[| E_{\nu}^{P} \rangle \langle  E_{\nu}^{P} |\right] =  \partial_{\phi}\bigg[ \sum_{\nu} | E_{\nu}^{P} \rangle \langle  E_{\nu}^{P} |\bigg] = \partial_{\phi}\mathbb{I} = 0 \text{.}
\end{align}
The second term can be easily evaluated:
\begin{align}
	\sum_{\nu} \frac{\langle E_{\nu}^{0} |c^\dagger_{j',\sigma'}\partial_{\phi} \left[| GS^P \rangle\langle GS^P |\right]\big|_{\phi=0} c_{j,\sigma}| E_{\nu}^{0} \rangle}{ \omega -U} = \partial_\phi \frac{\langle GS^P |c_{j,\sigma} c^\dagger_{j',\sigma'}| GS^P \rangle}{ \omega - U} \big|_{\phi=0} \text{.}
\end{align}
While the evaluation of the third term is more subtle and involves general results from degenerated perturbation theory \cite{Sakurai_2020}:
\begin{align}
	D\partial_{\phi} \langle  E_{\nu}^P  \vert H_H \vert E_{\nu}^P  \rangle \Big|_{\phi=0} &=  U \langle  E_{\nu}^0  \vert H_0 \vert E_{\nu}^0  \rangle \\
	H_0 \vert E_{\nu,l}^{0,m}  \rangle &= \langle E_{\nu,l}^{0,m} \vert H_0 \vert E_{\nu,l}^{0,m}  \rangle \vert E_{\nu,l}^{0,m}  \rangle + \sum_{\nu',l',m'; l'\neq l \wedge m'\neq m} c_{\nu',l',m'} \vert E_{\nu',l'}^{0,m'}  \rangle  \text{.}
\end{align}
Consequently we find:
\begin{align}
	-  &\sum_{\nu} \frac{\langle GS^0 |c_{j,\sigma}| E_{\nu}^{0} \rangle\langle E_{\nu}^0 |c^\dagger_{j',\sigma'}| GS^0 \rangle}{( \omega -U)^2}  \partial_{\phi} \left[E_0^P - E_{\nu}^{P}\right]\Big|_{\phi=0} \\ 
	&= -   U\frac{\langle GS^0 |c_{j,\sigma} c^\dagger_{j',\sigma'}| GS^0 \rangle\langle GS^0 |  H_0 | GS^0 \rangle }{D( \omega -U)^2}  +  U\frac{\langle GS^0 |c_{j,\sigma}  H_0 c^\dagger_{j',\sigma'}| GS^0 \rangle}{D( \omega -U)^2} \text{.} 
\end{align}
As the next step we take the expansion at low energy $\omega=0$ to the first order:
\begin{align}
	\mathcal{A}(\phi,\omega) \approx \mathcal{A}\big|_{\phi=0,\omega=0} +  \phi \partial_{\phi}\mathcal{A}\big|_{\phi=0,\omega=0} +  \omega \partial_{\omega}\mathcal{A}\big|_{\phi=0,\omega=0} \text{.}
\end{align}
The individual terms are:
\begin{align}
	\mathcal{A}\big|_{\phi=0,\omega=0} &= -\frac{1}{U} \langle GS^0 |c_{j,\sigma}c^\dagger_{j',\sigma'}| GS^0 \rangle \\
	\partial_{\phi}\mathcal{A}\big|_{\phi=0,\omega=0} &= -   \frac{1}{UD} \langle GS^0 |c_{j,\sigma} c^\dagger_{j',\sigma'}| GS^0 \rangle\langle GS^0 |  H_0 | GS^0 \rangle   +  \frac{1}{UD}\langle GS^0 |c_{j,\sigma}  H_0 c^\dagger_{j',\sigma'}| GS^0 \rangle \\
	&\quad \quad - \frac{1}{U} \partial_\phi \langle GS^P |c_{j,\sigma} c^\dagger_{j',\sigma'}| GS^P \rangle \big|_{\phi=0} \\
	\partial_{\omega}\mathcal{A}\big|_{\phi=0,\omega=0} &= - \frac{1}{U^2} \langle GS^0 |c_{j,\sigma}c^\dagger_{j',\sigma'}| GS^0 \rangle 
\end{align}
An equivalent derivation gives the result on $\mathcal{B}$, combining both results in the low energy Green's function approximation leads to:
\begin{align}
	G_{j,\sigma;j',\sigma'}( \omega) &\approx \frac{1}{U^2} \left[ H^{\text{z}}_{j,\sigma;j',\sigma'} - \omega \delta_{j,\sigma;j',\sigma'} \right] \\
	H^{\text{z}}_{j,\sigma;j',\sigma'} &=  -U \langle GS^0 |\left[c_{j,\sigma} , c^\dagger_{j',\sigma'} \right]| GS^0 \rangle - D\partial_\phi \langle GS^P |\left[c_{j,\sigma} , c^\dagger_{j',\sigma'} \right]| GS^P \rangle \big|_{\phi=0} \\ 
	&-  \langle GS^0 |\left[c_{j,\sigma} , c^\dagger_{j',\sigma'} \right]| GS^0 \rangle \langle GS^0 | H_0 | GS^0 \rangle +  \langle GS^0 |c_{j,\sigma}  H_0 c^\dagger_{j',\sigma'}| GS^0 \rangle \\ 
	&- \langle GS^0 |c^\dagger_{j',\sigma'}   H_0  c_{j,\sigma} | GS^0 \rangle \label{SM:SU2B_Hz}
\end{align}
At this stage we already derived one of our main results of the manuscript. We only assumed that the Hubbard Hamiltonian at study is strongly correlated $\phi\ll 1$ and at half-filling. 
In comparison with Eq.~\eqref{EQ:sm_gf_hubbard} we recognize almost all terms, up to the derivative $\partial_\phi \langle GS^P |\left[c_{j,\sigma} , c^\dagger_{j',\sigma'} \right]| GS^P \rangle \big|_{\phi=0}$, which can be treated by perturbation theory \cite{Zwiebach_2018}. 
In the case of an SU(2)-symmetric total Hamiltonian and an SU(2)-symmetric ground state, our expansion in Eq.~\eqref{EQ:sm_expansion} is sufficient and yields the exact result. This, however, is by no means trivial; in fact, it follows from second-order degenerate perturbation theory in combination with the spin-operator representation Eq.~\eqref{EQ:spin_rep} and the Wigner--Eckart theorem\cite{Sakurai_2020}, which allows one to significantly simplify the expectation value $\left. \partial_\phi \langle GS^P | \left[c_{j,\sigma}, c^\dagger_{j',\sigma'} \right] | GS^P \rangle \right|_{\phi=0}
$.

Morover, the \gls{gfz}-Hamiltonian in Eq.~\eqref{SM:SU2B_Hz} reveals already the structure of the \gls{gfz}-dispersion. In particular we recognize two different energy scales, $\mathcal{O}(U)$ and  $\mathcal{O}(D)$. The dominant zeroth order contribution $2U \Big[\langle GS^0 | \overrightarrow{S}_j  | GS^0 \rangle  \cdot \overrightarrow{\tau}\Big]_{\sigma,\sigma'} \delta_{j,j'}$, is local and directly related to the magnetic polarization of the strongly correlated groundstate. It shifts the zeros towards the Hubbard bands, what represents the main effect of a magnetic field.
The investigated fine-structure of the dispersion is created by the remaining terms of order $\mathcal{O}(D)$, its evaluation involves an exact treatment of the correlators, which can become very lengthy for the SU(2)-breaking states or/and Hamiltonians. Nonetheless, for the SU(2)-symmetric case we arrive at an very elegant results as we derive in Section~\ref{SM:GFA_SU2} and \ref{SM:rWGFA_SU2}.

\subsection{Self-consistent selfenergy construction}
The selfenergy $\Sigma$ is formally defined by the Dyson equation:
\begin{align}
	\Sigma = G_0^{-1} -  G^{-1} \text{.}
	\label{EQ:sm_dyson_zeros}
\end{align}
Here, the \gls{gfz} of $G$ imply poles of the selfenergy $\Sigma$, since the bare inverse \gls{gf} $G_0^{-1}$ is bounded.
Knowing the expansion of $G$ in terms of the correlation strength $\phi$, Eq.~\eqref{EQ:sm_dyson_zeros} implies a similar expansion for the selfenergy $\Sigma$, where we can eventually read out order-by-order an approximated selfenergy $\Sigma$. However, we have found that in general such an approximation does not preserve the \gls{gfz}.
Therefore, in the following we will construct a selfenergy which, for a given approximated \gls{gf} $G$, preserves the zeros and formally satisfies the Dyson equation. 
First, we factorize the roots of the \gls{gf} with an appropriate $H^z(\omega ,\phi)$:
\begin{align}
	G &= \mathcal{G}_{\text{H}} \tilde{G}^{-1} \\
	\tilde{G} &=   \left(\omega - H^z(\omega ,\phi) \right)^{-1} \text{,}  
\end{align}
here $\tilde{G}$ should correctly reproduce $G$ at its roots, but is not unique away from the roots.
The resulting $\mathcal{G}_{\text{H}}$ can be assumed to be locally invertible. 
Now we propose an ansatz for the selfenergy $\Sigma$ with an invertible $\mathcal{S}$:
\begin{align}
	\Sigma &= \left(\omega -  H^z\right)^{-1} \mathcal{S} = \tilde{G} \mathcal{S} \text{,}
\end{align}
which by construction leads to the correct poles. 
Now we find for the Dyson equation:
\begin{align}
	-G\Sigma =  -\mathcal{G}_{\text{H}} \mathcal{S} = \left[\mathbb{I} - \Sigma^{-1}G_0^{-1}  \right]^{-1} = \left[\mathbb{I} - \mathcal{S}^{-1}\tilde{G}^{-1}G_0^{-1}  \right]^{-1} \text{.}
	\label{EQ:sm_dyson_zeros2}
\end{align}
Assuming a small $\vert \vert \mathcal{S}^{-1}\tilde{G}^{-1}G_0^{-1} \vert \vert < 1 $ (self-consistency condition), we can expand the right site as a Neumann series and find:
\begin{align}
	\mathcal{G}_{\text{H}} \mathcal{S} = -\sum_{k=0} \left[\mathcal{S}^{-1}  \tilde{G}^{-1} G_0^{-1}\right]^k \text{.} 
\end{align}
As last step we multiply by $\mathcal{S}^{-1}$ from right and expand both ,$\mathcal{G}_{\text{H}}$ and $\mathcal{S}^{-1}$, in orders of $\phi$ (up to the highest known order $n$ of $\mathcal{G}_{\text{H}}$ ):
\begin{align}
	\mathcal{G}_{\text{H}}  &\approx \sum_{k=0}^n \phi^k \mathcal{G}_{\text{H},k} + \mathcal{O}(\phi^{n+1}) \text{ ,}  \\
	\mathcal{S}^{-1}  &\approx \sum_{k=0}^n \phi^k \mathcal{S}^{-1}_k + \mathcal{O}(\phi^{n+1}) \text{ ,}  \\
	\mathcal{G}_{\text{H}} 
	&\approx  -\sum_{k=0}^n \left[\left(\sum_{m=0}^n \phi^m \mathcal{S}^{-1}_m \right)  \tilde{G}^{-1} G_0^{-1}\right]^k \left(\sum_{l=0}^n \phi^l \mathcal{S}^{-1}_l \right) + \mathcal{O}(\phi^{n+1}) \text{.} 
	\label{EQ:sm_selfenergy_series}
\end{align}
Eq.~\eqref{EQ:sm_selfenergy_series} is valid for a vanishing zeroth order $\mathcal{G}_{\text{H},0} =0$, what implies $\mathcal{S}^{-1}_0=0$.  We then find an iterative series for $\mathcal{S}^{-1}_k$:
\begin{align}
	\mathcal{S}^{-1}_k =  -\mathcal{G}_{\text{H},k} - F_k(\mathcal{S}^{-1}_{k-1},...,\mathcal{S}^{-1}_{2}) \text{.}
\end{align}
The $F_k(\mathcal{S}^{-1}_{k-1},...,\mathcal{S}^{-1}_{2})$ are in general quite complicated, but do not involve any critical matrix inversion. Given all the $\mathcal{S}^{-1}_k$, we are able to find an approximation of $\mathcal{S}^{-1}$ and get a selfenergy $\Sigma = \tilde{G} \mathcal{S}$.
The constructed selfenergy $\Sigma $ does correctly reproduce the \gls{gf} $G$  and fulfills the Dyson equation up to the highest known order $n$ of $G$:
\begin{align}
	\norm{\mathcal{G}_{\text{H}} \tilde{G}^{-1} - \left[ G_0^{-1} - \Sigma \right]^{-1}}  &= \norm{ \mathcal{G}_{\text{H}} \tilde{G}^{-1} - \left[ \mathcal{S}^{-1} \tilde{G}^{-1} G_0^{-1} - \mathbb{I} \right]^{-1} \mathcal{S}^{-1} \tilde{G}^{-1} } \nonumber \\
	&\leq \norm{ \mathcal{G}_{\text{H}} - \left[ \mathcal{S}^{-1} \tilde{G}^{-1} G_0^{-1} - \mathbb{I} \right]^{-1}\mathcal{S}^{-1}} \norm{\tilde{G}^{-1}} \nonumber \\
	&\leq \norm{ \mathcal{G}_{\text{H}} + \sum_{k=0} \left[\mathcal{S}^{-1}  \tilde{G}^{-1} G_0^{-1}\right]^k  \mathcal{S}^{-1}} \norm{\tilde{G}^{-1}} \nonumber \\
	&\leq \norm{\sum_{k=n+1} \left[\left(\sum_{m=n+1} \phi^m \mathcal{S}^{-1}_m \right)  \tilde{G}^{-1} G_0^{-1}\right]^k \left(\sum_{l=n+1} \phi^l \mathcal{S}^{-1}_l \right)} \norm{\tilde{G}^{-1}} \nonumber \\
	&\leq \mathcal{O}(\phi^{n+1}) \norm{\tilde{G}^{-1}} = \mathcal{O}(\phi^{n+1})
\end{align}

\begin{figure}[htp]
	\includegraphics[width=0.5\linewidth]{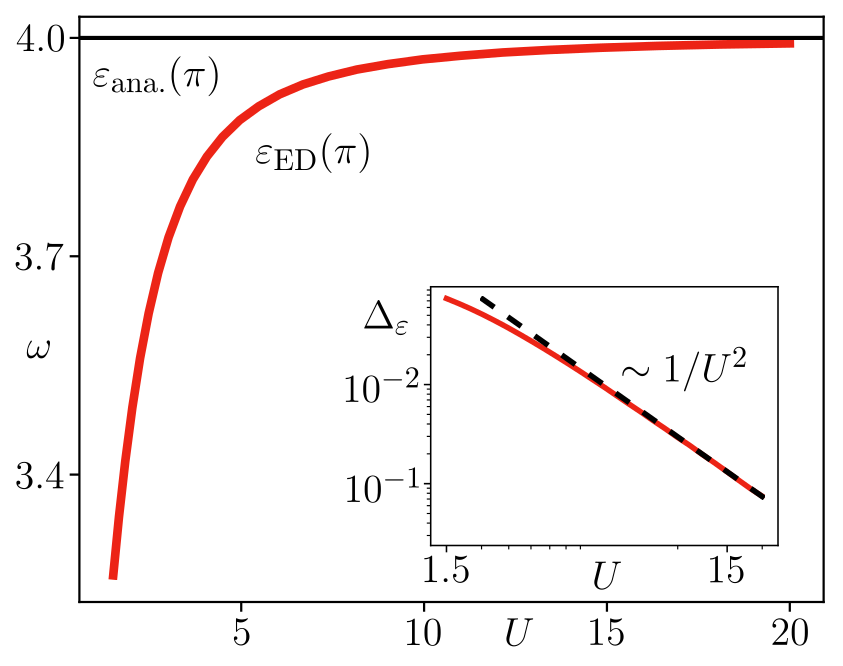}
	\caption{Convergence of the \gls{gfz}-dispersion $\varepsilon_{\text{ED}}$ (black) against our analytical prediction in the strongly correlated limit $\varepsilon_{\text{ana}}$(red), exemplarily for $k=\pi$. SSH-Hubbard chain ($N=8$, $w = v = 1.0$, PBC, ED).
		Inset: distance $\Delta_{\varepsilon} = \vert \varepsilon_{\text{ED}} - \varepsilon_{\text{ana}} \vert $ and fit (dashed).
	}
	\label{fig:gfz_distance}
\end{figure}

As an minimal example we determine the selfenergy corresponding to the previously derived \gls{gf} $G$ in Eq.~\eqref{EQ:sm_gf_hubbard}:
\begin{align}
	G(\omega,\phi) \approx  \frac{1}{D^2} \phi^2 \left[ H^z - \omega \right] + \mathcal{O}(\phi^3)
\end{align}
a natural decomposition is:
\begin{align}
	\tilde{G}^{-1}  = -\left[ H^z - \omega  \right]  \quad \text{ and }\quad \mathcal{G}_{\text{H}} = -\frac{\phi^2}{D^2} \mathbb{I} \text{.} \nonumber
\end{align}
Order-by-order, Eq.~\eqref{EQ:sm_selfenergy_series} then leads to:
\begin{align}
	\mathcal{S}^{-1}_0 = 0 \quad
	\mathcal{S}^{-1}_1 = 0 \quad
	\mathcal{S}^{-1}_2 =  \frac{1}{D^2} \mathbb{I} \text{.} \nonumber
\end{align}
Eventually we can approximate the selfenergy by:
\begin{align}
	\Sigma  &\approx \frac{D^2}{\phi^2} \left[\omega -H^z  \right]^{-1}  \text{,}
\end{align}
what is inline with \cite{Wagner_2023}.
\updt{
	\subsection{Convergence of the Green's Function Zero Dispersion}
	Our low-energy \gls{gf}-approximation is valid in the strongly correlated limit. To probe the convergence of the \gls{gfz}-dispersion against our analytical prediction, we computed the exact \gls{gfz}s using \gls{ed} for increasing interaction strengths $U$. In Fig.~\ref{fig:gfz_distance}, we present the exact frequency of the \gls{gfz} for a fixed momentum $k=\pi$. As can be clearly seen, the exact results converge rapidly to our prediction with increasing correlation strength, as expected. Surprisingly, even at a relatively weak interaction strength of $U=5$, the absolute error for the given parameters is already of order $\sim 10^{-1}$. In the inset, we show the mean distance between the two dispersions and find, on a double-logarithmic scale, the expected $\sim 1/U^2$ scaling of the absolute error.}

\section{Perturbation theory for two additional lead quantum dots}
In the following, we investigate the co-tunneling process through an $SU(2)$-symmetric Hubbard model. In addition to the Hubbard Hamiltonian $H_{\text{H}}$ from Eq.~\eqref{EQ:sm_hubbard_ham}, we introduce two auxiliary quantum dots described by $H_{\text{D}}$, with the two subsystems being weakly coupled via $C$. The total Hamiltonian is then given by:

\begin{align} 
	H_{\text{tot}} &= H_{\text{H}} + \gamma C + H_{\text{D}}, \label{EQ:sm_htot}
	\\ C &= \sum_{j,\sigma} l_j c^\dagger_{j\sigma} c_{L,\sigma} + r_j c^\dagger_{j\sigma} c_{R,\sigma} + \text{h.c.} ,
	\\ H_{\text{D}} &= \sum_{\sigma} \mu_{\text{L}}c_{L,\sigma}^\dagger c_{L,\sigma} + \mu_{\text{R}}c_{R,\sigma}^\dagger c_{R,\sigma} + 2U \sum_{s\in{L,R}} n_{s,\uparrow}n_{s,\downarrow}. \label{EQ:sm_exp_tot_H} \end{align}

Here, the small dimensionless parameter $\gamma$ quantifies the coupling between the system and the quantum dots, while the tunneling amplitudes $l_j$ and $r_j$ describe the local coupling structure, which can be chosen arbitrarily. The total Hamiltonian $H_{\text{tot}}$ is assumed to exhibit $SU(2)$-symmetry. The auxiliary dots are detuned by $U$ from the central Hubbard system $H_{\text{H}}$, ensuring that an additional particle beyond the half-filled Hubbard system is forced onto the lead dots. In the following, we focus on this regime, where the Hubbard system remains half-filled, and a single additional particle occupies the quantum dots. Further fine-tuning of the dot potentials is described by the weak $\mu_{\text{L}}$, $\mu_{\text{R}}$. For the unperturbed system, where $\gamma = 0$, and weak dot potentials, $\mu_L$ and $\mu_R$, we obtain four low-energy states that describe the low-energy subspace $\mathcal{H}^{0}$:

\begin{align}
	\mathcal{H}^{0} &= \mathcal{H}^{D} \otimes 	\vert GS \rangle  = \{ \vert s,\sigma, GS \rangle \}~ \text{with } s\in\{ L,R \},\sigma\in\{\uparrow,\downarrow\}  \\
	\mathcal{H}^{D} &= \{\vert 0,\uparrow \rangle, \vert 0,\downarrow \rangle,\vert \uparrow,0 \rangle, \vert \downarrow,0 \rangle\} 
\end{align}

Here, the additional electron, with spin $\sigma \in \{\uparrow, \downarrow\}$, can occupy either the left or right quantum dot, labeled by $s \in \{L, R\}$, while the central Mott insulator is always in its half-filled groundstate.

We now apply a Schrieffer-Wolff transformation \cite{Schrieffer_1966, Bir_1974, Bravyi_2011,Ripoll_2022}, projecting $H_{\text{tot}}$ on the low-energy states $\mathcal{H}^{0}$. The tunneling term $\gamma C$ is treated as a perturbation. The resulting effective Hamiltonian $\bar{h}$ will describe the co-tunneling process at zero temperature and, up to second order in $\gamma$, is given by (see, e.g. Eq. A36 in~\cite{Ripoll_2022}):
\begin{align}
	\bar{h}_{s,\sigma;s',\sigma'} &\approx E_{GS} + \langle  s,\sigma \vert H_{\text{D}} \vert s',\sigma'\rangle + \frac{\gamma^2}{2}\sum_{k} \frac{\langle  s,\sigma,GS \vert C \vert  E_{k} \rangle  \langle  E_{k} \vert   C \vert s',\sigma',GS \rangle}{ \mu_s  +  E_{GS} - E_k} \nonumber \\
	&\quad +  \frac{\gamma^2}{2}\sum_{k} \frac{\langle  s,\sigma,GS \vert C \vert  E_{k} \rangle  \langle  E_{k} \vert   C \vert s',\sigma',GS \rangle}{ \mu_{s'}  +  E_{GS} - E_k} + \mathcal{O}(\gamma^3) \text{.} \label{EQ:sm_h_eff_general}
\end{align}
Here, we sum over all eigenstates $E_k$ that are complementary to the low-energy subspace $\mathcal{H}^{0}$. A careful evaluation of Eq.~\eqref{EQ:sm_h_eff_general} then leads to:
\begin{align}
	\bar{h} &= \begin{pmatrix}
		\left(\mu_{\text{L}}-\gamma^2\sum_\sigma \langle  r \sigma \vert G^B(\mu_{R}) \vert r \sigma \rangle \right)\tau_0 + \gamma^2  h_l(\mu_{\text{L}})  & \frac{\gamma^2}{2} \left[ h_t^\dagger(\mu_{\text{L}}) + h_t^\dagger(\mu_{\text{R}})\right]  \\
		\frac{\gamma^2}{2} \left[ h_t(\mu_{\text{L}}) + h_t(\mu_{\text{R}})\right] & \left(\mu_{\text{R}}-\gamma^2\sum_\sigma \langle  l \sigma \vert G^B(\mu_{L}) \vert l \sigma \rangle \right)\tau_0 + \gamma^2  h_r(\mu_{\text{R}})
	\end{pmatrix}  \label{EQ:sm_heff_gform}\\
	h_{s\in\{l,r\}}(\omega) &= \begin{pmatrix}
		\langle  s\uparrow \vert G^A(\omega) \vert s\uparrow \rangle   -\langle  s \downarrow \vert G^B(\omega+2U) \vert s  \downarrow \rangle  & \langle  s\uparrow \vert G^A(\omega) \vert s\downarrow \rangle +  \langle  s \downarrow \vert G^B(\omega+2U) \vert s  \uparrow \rangle \\
		\langle  s\downarrow \vert G^A(\omega) \vert s\uparrow \rangle +  \langle  s \uparrow \vert G^B(\omega+2U) \vert s  \downarrow \rangle & \langle  s\downarrow \vert G^A(\omega) \vert s\downarrow \rangle  -\langle  s \uparrow \vert G^B(\omega+2U) \vert s  \uparrow \rangle
	\end{pmatrix} \\
	h_t(\omega) &= \begin{pmatrix}
		\langle  r\uparrow \vert G(\omega) \vert l\uparrow \rangle & \langle  r\uparrow \vert G(\omega) \vert l\downarrow \rangle \\
		\langle  r\downarrow \vert G(\omega) \vert l\uparrow \rangle & \langle  r\downarrow \vert G(\omega) \vert l\downarrow \rangle 
	\end{pmatrix}  
\end{align}
For brevity, we introduced the tunneling vectors $\vert s\sigma \rangle$, encoding the amplitudes $l_i,r_i$ by $\langle  i \sigma \vert l \sigma' \rangle = \delta_{\sigma,\sigma'} l_i$ (same for $\{r_i\}$). They act on the matrix-valued \gls{gf}s $G,G^A,G^B$. The \gls{gf} components are defined as: 
\begin{align}
	G_{j,\sigma ; j',\sigma'}(\omega) &=  G^A_{j,\sigma ; j',\sigma'}(\omega) + G^B_{j,\sigma ; j',\sigma'}(\omega) \\
	G^A_{j,\sigma ; j',\sigma'}(\omega) &= \sum_{\eta,l}  \frac{\langle  E_{GS} \vert    c_{j,\sigma}  \vert E^{+}_{\eta,l} \rangle \langle  E^{+}_{\eta,l} \vert    c^\dagger_{j',\sigma'}  \vert E_{GS} \rangle }{\omega +E_{GS} - E^{+}_{\eta,l}  }  \\
	G^B_{j,\sigma ; j',\sigma'}(\omega) &= \sum_{\eta,l} \frac{\langle  E_{GS} \vert   c^\dagger_{j',\sigma'} \vert E^{-}_{\eta,l} \rangle  \langle  E^{-}_{\eta,l} \vert   c_{j,\sigma}  \vert E_{GS} \rangle }{\omega  + E^{-}_{\eta,l}  - E_{GS}}
\end{align}
Remarkably, the microscopic details of the central Hubbard system in the effective Hamiltonian $\bar{h}$ are fully captured by the \gls{gf} $G$  and its constituents $G^A,G^B$. For large interaction strengths $U$, that will be a Mott insulator for which we already derived its \gls{gf} in the previous section. Following the same steps (Eq.~\eqref{EQ:sm_rewr_ons_a}, Eq.~\eqref{EQ:sm_rewr_ons_b}, Eq.~\eqref{EQ:sm_rewr_nn_t}, Eq.~\eqref{EQ:sm_rewr_nn_r} and Eq.~\eqref{EQ:sm_rewr_nn_w}) we derive the Mott-insulating \gls{gf} up to the third order in $\phi$. For a $SU(2)$-symmetric Hubbard system with a unique, non-magnetic groundstate we find:
\begin{align}
	G^A(\omega + \alpha U) &\approx \frac{1}{2U^2} \frac{1}{\left(\alpha-1\right)^2} \left[H^{\text{z}} -\omega -U +\alpha U + \frac{\alpha}{2} \left( H_0 - H^{\text{z}}\right) \right]  + \mathcal{O}(\phi^3) \\
	G^B(\omega + \alpha U ) &\approx \frac{1}{2U^2} \frac{1}{\left(\alpha+1\right)^2} \left[H^{\text{z}} -\omega +U +\alpha U - \frac{\alpha}{2} \left( H_0 - H^{\text{z}}\right) \right] + \mathcal{O}(\phi^3) \text{.}
\end{align}
The \gls{gfz}-Hamiltonian $H^{\text{z}}$ was already derived in Eq.~\eqref{EQ:sm_final_zh}, while $H_0$ denotes the bare non-interacting Hamiltonian. The result is valid for $\omega \ll U$, where either $\alpha=0$ or $\vert \alpha\vert \geq 2$. For $0 <\vert \alpha\vert < 2$ the expansion is close to the singularities in the Hubbard band. Together with Eq.~\eqref{EQ:sm_heff_gform} we can further simplify:
\begin{align}
	\bar{h} &= \begin{pmatrix}
		(\mu_{\text{L}} + a_L(\mu_{\text{L}}) - 2 b_R(\mu_{\text{R}}) )\tau_0  & t^\ast_{\text{co}}(\bar{\mu}) \tau_0  \\
		t_{\text{co}}(\bar{\mu}) \tau_0 & (\mu_{\text{R}} + a_R(\mu_{\text{R}}) - 2 b_L(\mu_{\text{L}}))\tau_0
	\end{pmatrix}  \label{EQ:heff_spin}\\
	a_{s\in \{ \text{L} , \text{R} \}}(\omega) & = a_{s\sigma}(\omega) = \frac{\gamma^2}{2U^2} \langle  s \sigma \vert \left[\frac{7}{9}H^{\text{z}} - \frac{8}{9} \omega - \frac{4}{3}U+ \frac{1}{9} H_0 \right] \vert s \sigma\rangle \\
	b_{s\in \{ \text{L} , \text{R} \}}(\omega) &= b_{s\sigma}(\omega) = \frac{\gamma^2}{2U^2} \langle  s \sigma \vert \left[H^{\text{z}} - \omega + U \right] \vert s \sigma \rangle  \\
	t_{\text{co}}(\bar{\mu}) &= t_{co,\sigma}(\bar{\mu}) = \frac{\gamma^2}{U^2} \langle  r \sigma \vert\left[H^{\text{z}} - \bar{\mu} \right] \vert l \sigma \rangle \quad \quad \bar{\mu} = \frac{1}{2} \left(\mu_{\text{L}} + \mu_{\text{R}}\right)
	\label{EQ:sm_eff_ham_bath} 
\end{align}
Notice that due to spin isotropy $a_{s\uparrow}=a_{s\downarrow}$, $b_{s\uparrow}=a_{b\downarrow}$ and $t_{co,\uparrow}=t_{co,\downarrow}$, thus for brevity reasons we can also discard $\tau_0$. The effective Hamiltonian $\bar{h}$ describes the co-tunneling dynamics through the Mott insulator at zero temperature and is fully defined on the dots. Single entries of $H^{\text{z}}$ can be accessed by choosing local tunnel couplings, i.e., $l_k \propto \delta_{ik},~ r_{k'}\propto \delta_{jk'}, ~i\neq j$ then $ t_{\text{co}}\propto \left(H^{\text{z}}\right)_{i,j}$. By fine-tuning the dot potentials, $\mu_{\text{L}}$ and $\mu_{\text{R}}$, it is also possible to make the diagonal entries in $\bar{h}$ vanish.

\section{Charge stability diagram}
We compute the stability diagram by determining the global groundstate of the full Hamiltonian $H_{\text{tot}}$ in Eq.~\eqref{EQ:sm_htot} for varying weak dot potentials $\mu_{\text{L}},\mu_{\text{R}}$.
The groundstate obeys a half-filled Hubbard system for small potentials $\mu_{\text{L}},\mu_{\text{R}}$, while there are either zero $(0,0)$, one $(1,0)$ (dominantly at the left dot), $(0,1)$ (dominantly at the right dot) or two $(1,1)$ particles on the auxiliary dots. We then determine the dot occupations $N_L(\mu_{\text{L}},\mu_{\text{R}}),N_R(\mu_{\text{L}},\mu_{\text{R}})$. With the onset of the co-tunneling induced coupling between the auxiliary dots, we find that there is a smooth transition between the two localized regimes $(1,0)$ and $(0,1)$. 
The resulting charge curve $N_L(\mu_{\text{L}},\mu_{\text{R}})$ ,$N_R(\mu_{\text{L}},\mu_{\text{R}})$ in this regime can be derived from  Eq.~\eqref{EQ:heff_spin}. Let us rewrite:
\begin{align}
	\bar{h} &=  C_0(\bar{\mu}, \delta \mu) \tau_0  + \begin{pmatrix}
		C_1\bar{\mu} + C_2 \delta \mu + C_3  & t^\ast_{\text{co}}(\bar{\mu})   \\
		t_{\text{co}}(\bar{\mu})  & -(C_1\bar{\mu} + C_2 \delta \mu + C_3)
	\end{pmatrix} \\
	\delta \mu &= \mu_{\text{L}} -\mu_{\text{R}} \quad \quad \bar{\mu} = \frac{1}{2} \left(\mu_{\text{L}} + \mu_{\text{R}}\right) 
\end{align}
Where we used that $a_s(\omega)$ and $b_s(\omega)$ are linear in the frequency $\omega$. Now for the eigenstates we can neglect the overall shift $C_0$ and for a fixed $\bar{\mu}$ we introduce the shifted potential difference $\tilde{\delta} \mu = \delta\mu + \frac{ C_1\bar{\mu}  + C_3}{C_2}$:
\begin{align}
	\bar{h} &=  \begin{pmatrix}
		C_2 \tilde{\delta}\mu  & t^\ast_{\text{co}}(\bar{\mu})   \\
		t_{\text{co}}(\bar{\mu})  & - C_2 \tilde{\delta}\mu
	\end{pmatrix} \\ \text{.}
\end{align}
The lowest eigenstate $\vert \varepsilon_- \rangle$ is:
\begin{align}
	\vert \varepsilon_- \rangle &= \frac{1}{\sqrt{\left( \chi - \sqrt{\chi^2 +1} \right)^2 + 1}} \begin{pmatrix}
		\chi - \sqrt{\chi^2 +1} \\
		\frac{ t_{\text{co}}(\bar{\mu}) }{\vert t_{\text{co}}(\bar{\mu}) \vert }
	\end{pmatrix} \\
	\chi &= \frac{C_2 \tilde{\delta}\mu}{ \vert t_{\text{co}}(\bar{\mu}) \vert }
\end{align}
The occupations on the dots $N_L(\bar{\mu},\tilde{\delta}\mu)$,$N_R(\bar{\mu},\tilde{\delta}\mu)$ are then given by:
\begin{align}
	N_R(\bar{\mu},\tilde{\delta}\mu) &= \frac{1}{\left( \chi - \sqrt{\chi^2 +1} \right)^2 + 1} \\
	N_L &= 1 - N_R
\end{align}
For a fixed $\bar{\mu}$ we calculate the derivative with respect to the bare $\delta \mu$:
\begin{align}
	\frac{d N_R(\bar{\mu},\tilde{\delta}\mu)}{d (\delta \mu)} &= \frac{1}{2} \frac{1}{\left( \sqrt{\chi^2 + 1} \right)^3} \frac{d \chi}{d (\delta \mu) } = \frac{1}{2} \frac{1}{\left( \sqrt{\chi^2 + 1} \right)^3} \frac{C_2}{ \vert t_{\text{co}}(\bar{\mu}) \vert} 
\end{align}
The Full-Width-at-Half-Maximum is then given by $\Delta_{FWHM}= 2 \sqrt{2^{\frac{2}{3}}-1} \frac{\vert t_{\text{co}}(\bar{\mu} )\vert }{C_2}  \approx 1.53 \frac{\vert t_{\text{co}}(\bar{\mu} )\vert }{C_2} $. Moreover, we can approximate $C_2 \approx \frac{1}{2} +\mathcal{O}(\frac{\gamma^2}{U^2})$ and in the case of a strictly local system-dot couplings $l_k \propto \delta_{ik}, r_{k'}\propto \delta_{jk'}, i\neq j$, the co-tunneling $t_{\text{co}}$ is  also independent of $\bar{\mu}$. In total, at zero temperature, we find a universal curve with $\Delta_{FWHM}\approx 3.07 \vert t_{\text{co}} \vert $.

\section{Deviations beyond our perturbative regime for state-of-the-art quantum dot parameters}
\begin{figure}[htp]
	\includegraphics[width=1.0\linewidth]{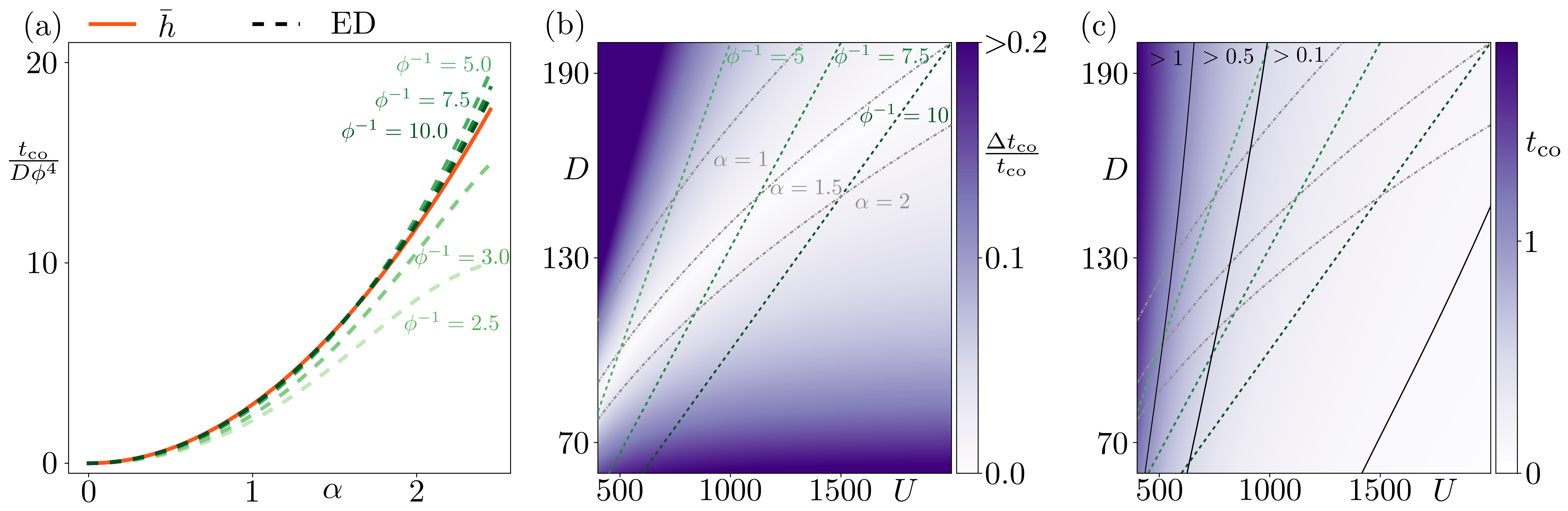}
	\caption{Co-tunneling $t_{\text{co}}$: numerical results (ED) vs analytical results (Eq.~\eqref{EQ:sm_eff_ham_bath}) for the SSH-Hubbard chain ($N=16$, $w = 0.8v$, PBC, $l_i = \delta_{i,1} $,$r_i = \delta_{i,2} $). In (a): Co-tunneling $t_{\text{co}}$ as function of $\alpha$ and different relative interaction strengths $\phi$, $D=100$ fixed, numerical results(dashed, green) and analytical results(solid,red). In (b): relative error of $t_{\text{co}}$ determined by Eq.~\eqref{EQ:sm_eff_ham_bath} compared to the exact result for various $D$,$U$ and fixed $\gamma=30$. In (c): absolute value of $t_{\text{co}}$, same parameters as in (b). As reference: contour lines of constant $\alpha$, $\phi$ and $t_{\text{co}}$  (grey, green,black).}
	\label{fig:exp_parameters}
\end{figure}
In the literature, a wide range of realistic quantum dot parameters  $\gamma,U$ and $D$ can be found~\cite{Barthelemy_2013,Schleser_2005,Braakman_2013,Sanchez_2014,Bordin_2024,Wang_2023,Ranni_2021,Schindele_2014,Liu_2022} . Accordingly, in Fig.~\ref{fig:exp_parameters}, we present an overview of the parameter regime where our perturbative approach is applicable and estimate the expected relative error, which accounts for the impact of neglected higher-order perturbations. In the derivation of Eq.~\eqref{EQ:heff_spin}, we assumed the limit of large interactions ($U$) and weak lead-system coupling ($\gamma$), a regime where the co-tunneling amplitude $t_{\text{co}}$ is expected to be a small quantity of order $\mathcal{O}(\frac{\gamma^2}{U^2})$. While this presents a potential limitation, we find that macroscopic values of $t_{\text{co}}$ can still be achieved with sufficient accuracy within our theoretical framework.
The three material parameters $\gamma,U$ and $D$ can be re-parametrized as:
\begin{align}
	\phi^{-1} = \frac{U}{D}, \quad \alpha =  \frac{\gamma U}{D^2} \text{.}
\end{align}
Both are sufficient to describe the universal behavior, as a global rescaling of the original parameters does not affect the underlying physics. 
The previously discussed parameter $\phi$ characterizes the Mottness of the central Hubbard system, while $\alpha$ represents the ratio of $\gamma$ to the well-known exchange coupling $J = \frac{D^2}{U}$, which is related to the half-bandwidth of the effective low-energy Heisenberg model.
In Fig.~\ref{fig:exp_parameters}(a), we present the co-tunneling amplitude $t_{\text{co}}$ as function of the coupling ratio $\alpha$ and for various relative interaction strengths $\phi$ in the SSH-Hubbard chain ($w=0.8v$, PBC, $D=(v+w)\slash 2$) with local couplings $l_i = \delta_{i,1} $ and $r_i = \delta_{i,2} $:
\begin{align}
	H_0 = \sum_{j,\sigma} v c_{2j,\sigma}^\dagger c_{2j-1,\sigma} + w c_{2j+1,\sigma}^\dagger c_{2j,\sigma} + \text{h.c. .}
\end{align}
We find that the exact results (dashed, green) show excellent agreement with our analytical approach (solid, red), as given by Eq.~\eqref{EQ:heff_spin}, already at $\phi\approx 0.2 $ and up to relatively large values of 
$\alpha\approx 2$. As expected, increasing Mottness and weaken the lead-system coupling cause the exact results to converge toward the analytical predictions. More surprising, however, is the large value $\alpha$ with a good agreement, which we interpret as effect of the  vanishing first-order contributions of the weak lead-system hopping $C$ in Eq.~\eqref{EQ:sm_exp_tot_H}. The drastic change for $\phi>0.2$ signals the Mott transition. From Fig.~\ref{fig:exp_parameters}(a), the bare parameters $\gamma,U$ and $D$ can be directly extracted. For example,  for $\phi=0.2$ and $\alpha=1.5$ let us fix a half-bandwidth $D=100 \mu eV$, then we obtain $U=500\mu eV$, $\gamma=30\mu eV$, and $t_{\text{co}} \approx 1 \mu eV$.

In Fig.~\ref{fig:exp_parameters} (b), we present the relative error:
\begin{align}
	\frac{\Delta t_{\text{co}}}{t_{\text{co}}}=\frac{t_{\text{co}}^{\text{ana}} - t_{\text{co}}^{\text{exact}} }{t_{\text{co}}^{\text{ana}}}
\end{align}
as a function of the bare parameters $U$ and $D$,  with a fixed lead-system coupling $\gamma=30$. For reference, we also include contour lines corresponding to constant values of 
$\phi$ and $\alpha$. The relative error $\frac{\Delta t_{\text{co}}}{t_{\text{co}}}$ does show the relevance of the neglected higher order contribution in our theoretical framework.
As shown in Figure~\ref{fig:exp_parameters}~(b) there is a broad parameter regime where the relative error remains sufficiently small. However, notable deviations occur in two limiting cases: (i) for large half-bandwidth 
$D$, where the Mott transition takes place, and (ii) for strong lead-system coupling $\gamma$ relative to $D$.
The expected co-tunneling amplitudes $t_{\text{co}}$ in this regime are shown in Fig.~\ref{fig:exp_parameters} (c). A macroscopic co-tunneling amplitude $t_{\text{co}} > 1 \mu eV$ can be reached at smaller interaction strengths $U$, while this is also tied to a larger relative error (see Fig.~\ref{fig:exp_parameters} (b)). Notably, all results presented here exhibit scale invariance: a global scaling of $U$,$D$ and $\gamma$ by a factor $\lambda$ results in a corresponding scaling of the co-tunneling amplitude, $t_{\text{co}} \rightarrow \lambda t_{\text{co}}$, while the overall structure of the figures remains unchanged, apart from the axes rescaling.

\section{Renormalization of the GFZ-Hamiltonian by the spin-spin correlation functions}
\begin{figure}[htp]
	\centering
	\includegraphics[width=0.25\linewidth]{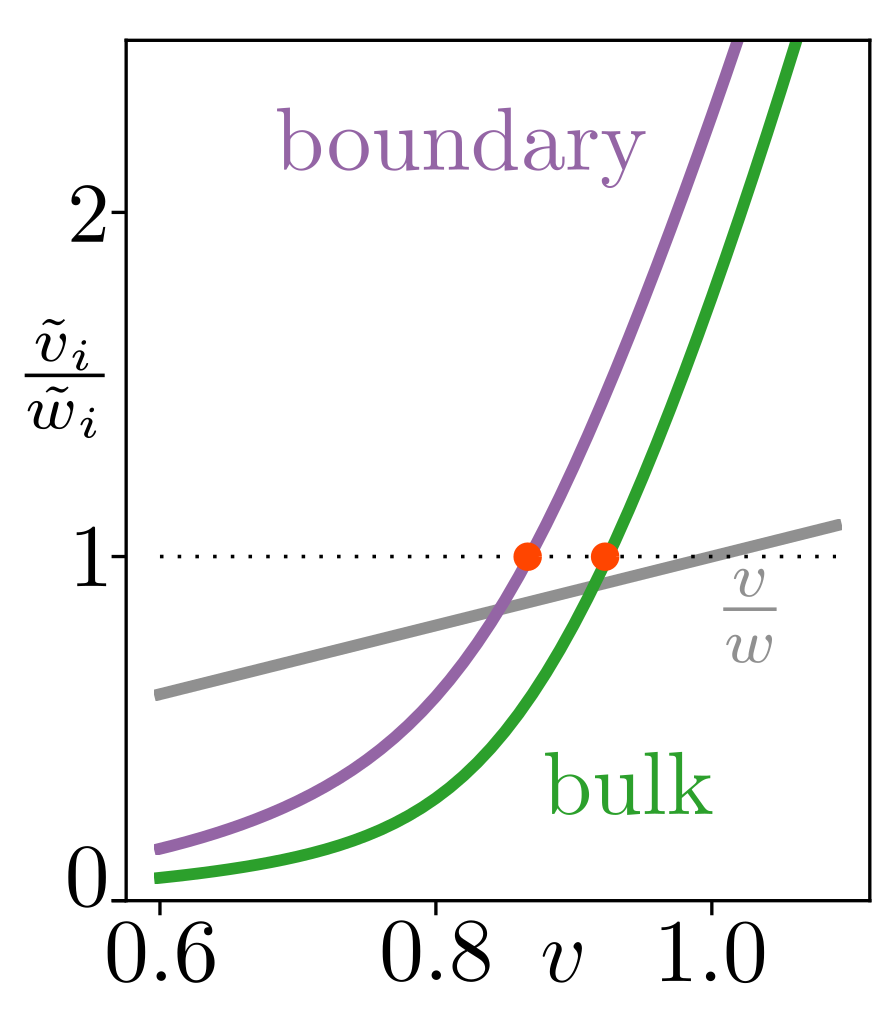}
	\caption{Staggering for OBC inverting at the boundary close to topological transition $v =w$. Here:\gls{ed},$U=200$ $N=20$, $\gamma=2$, $w = 20$, OBC), with $l_i = \delta_{i,1} $,$r_i = \delta_{i,2} $ (violet) and $l_i = \delta_{i,3} $,$r_i = \delta_{i,4} $ (green).}
	\label{fig:sm_osc_bbc}
\end{figure}
We comment on the renormalization of the \gls{gfz}-Hamiltonian of Eq.~\eqref{EQ:sm_final_zh} by the spin-spin correlation functions. In particular, we focus on the topological features of the underlying SSH model. We consider a finite-size SSH-Hubbard chain, where the non-interacting system shows zero-energy edge modes as a function of the staggered hopping ratio $v/w$.
As described by to Eq.~\eqref{EQ:sm_final_zh}, the bare hoppings get renormalized significantly by the spin-correlation values. Since the \gls{ssh} model at large U is effectively a spin chain, its ground state will obey \gls{aklt} physics \cite{Affleck_1987}, with the formation of spin-singlet dimers in correspondence with the stronger bonds. As a consequence, the original imbalance between the non-interacting $v$ and $w$ is further amplified.
In Fig.\eqref{fig:sm_osc_bbc} we show the behavior of the renormalized ratio $\frac{\tilde{v}}{\tilde{w}}$ as a function of $v$, for $w=20$ and $U=200$. The grey line shows the behavior of the bare hopping ratio. It is immediate to notice how the renormalized ratio varies much more heavily, passing from strong suppression to strong enhancement. The purple and green curves in the plot refer to different positions along the finite \gls{ssh} chain, respectively near the boundary and deep in the bulk. Finite-size effect entail a different renormalization in the two cases, and the topological transition, denoted by the red dots, happens at the boundary sites for a lower value of $v/w$ than in the bulk. In the region between the two red dots, the boundary sites are already in the trivial phase, while the bulk is topological. As a consequence, edge state reconstruction ensues \cite{Amaricci_2017}, with the topological edge zero described in \cite{Wagner_2023} shifting into the bulk.

\section{Minimal chain to resolve the renormalization}
\begin{figure}[htp]
	\centering
	\includegraphics[width=1.0\linewidth]{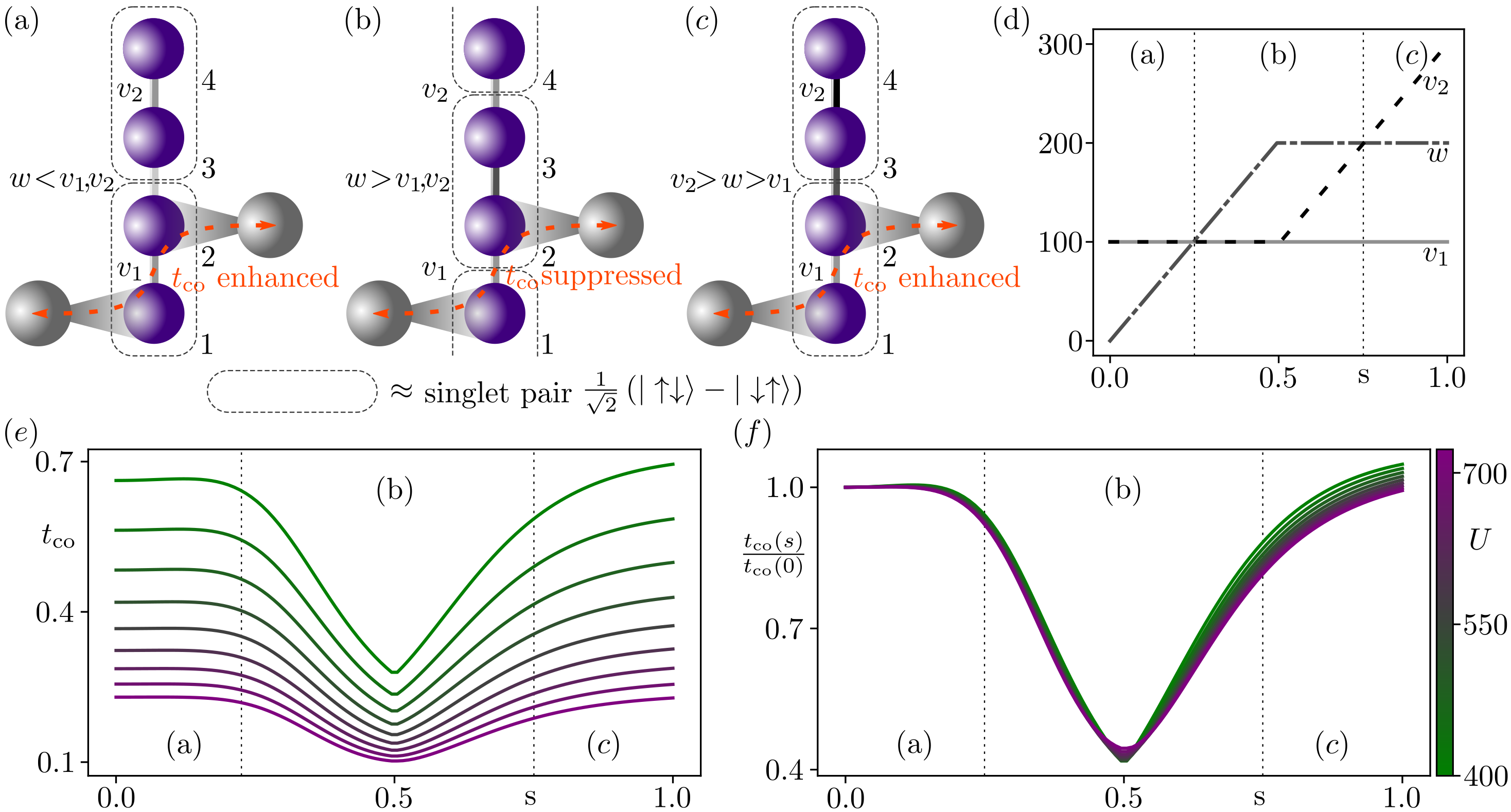}
	\caption{In (a),(b),(c): small chain of six quantum dots (four as Hubbard system, two as leads). The lead dots are at site 1 and 2 with $l_i = \delta_{i,1} $,$r_i = \delta_{i,2} $. In (a): Two weakly coupled dimers, groundstate approximately described by spin-singlets at bonds 1-2 and 3-4, the probed $t_{\text{co}}$ at bond 1-2 is enhanced by the spin correlation $\langle  \overrightarrow{S}_1 \cdot \overrightarrow{S}_2 \rangle $. In (b): groundstate resembles spin-singlet at bonds 2-3 and 1-4, spin at site 1 and 2 are roughly uncorrelated and probed $t_{\text{co}}$ at bond 1-2 is suppressed. In (c): large coupling at bond 3-4 restoring initial singlet structure, spins at site 1 and 2 are correlated again and $t_{\text{co}}$ is enhanced. In (d): parameter trace for which we computed the expected $t_{\text{co}}$, passing through all three regimes. In (e) and (f): expected (normalized) co-tunneling $t_{\text{co}}$ as function of $s$ and $U$ (\gls{ed},  $N=12$, fixed $U=400$ on the lead dots, $v_1=100$, $l_i = \delta_{i,1} $,$r_i = \delta_{i,2} $, $\gamma=20$).}
	\label{fig:sm_bcd}
\end{figure}
The minimal chain to resolve the the renormalization by the spin-spin correlators consists of four quantum dots modeling the central Hubbard system. Thereby, they are coupled pairwise by the tunable tunnelings $w,v_1,v_2$ (see Fig.\ref{fig:sm_bcd} (a), (b), (c)). 
Then, by varying $w$ and $v_2$ ($v_1$ is fixed), we probe three regimes: first for $v_1,v_2 >w$ (Fig.\ref{fig:sm_bcd} (a)), the system behaves as two uncoupled Hubbard dimers. In the strongly correlated limit its groundstate is then approximately described by two spin-singlets along the outer bonds (1-2, 3-4). Secondly, for $w>v_1,v_2$ (Fig.\ref{fig:sm_bcd} (b)), the situation is quite different, now we find to a good approximation a spin-singlet at the central bond (2-3) and a singlet between the spins at the edges (1-4). In the last case (Fig.\ref{fig:sm_bcd} (b)),  $v_2 >w> v_1$, there are again two spin-singlets along the outer bonds (1-2, 3-4) as in (a). 
As a result, the spin correlation along the probed, lower bond $\langle  \overrightarrow{S}_1 \cdot \overrightarrow{S}_2\rangle$ strongly varies for all three regimes. 
In the first case (a), both the spins at site 1 and site 2 are almost maximally entangled, the probed  spin correlation $\langle  \overrightarrow{S}_1 \cdot \overrightarrow{S}_2\rangle$ becomes large.
Whereas in the second case (b), $\langle  \overrightarrow{S}_1 \cdot \overrightarrow{S}_2\rangle$ is decreased, as both the spins at site 1 and site 2 are essentially uncorrelated. Furthermore, deep in the regime (c), the initial singlet pairs are restored and the spin correlation gets large again. As we found in the letter, via Eq.~\eqref{EQ:sm_final_zh} the correlation function significantly alters the hopping in the \gls{gfz} Hamiltonian $H^z$ at the probed bond and in consequence the measured co-tunneling amplitude $t_{\text{co}}$. Remarkably, in the presented setup Fig.\ref{fig:sm_bcd}, it is possible to tune the probed co-tunneling amplitude at bond 1-2 only by varying the couplings $w,v_2$ along the other bonds, while $v_1$ is fixed. This offers a way to probe the renormalization by the spin-spin correlation functions of the \gls{gfz} in an actual experiment by a clear qualitative signature.
To illustrate this, we computed the expected co-tunneling amplitude $t_{\text{co}}$ in Fig.\ref{fig:sm_bcd} (e) for all three regimes and different correlation strengths $U$, by probing a trace of various coupling parameters $w,v_1,v_2$ as shown in Fig.\ref{fig:sm_bcd} (d).
As predicted by Eq.~\eqref{EQ:sm_final_zh}, for the regime (a) ($v_1,v_2 >w$) the co-tunneling amplitude $t_{\text{co}}$ is large and we find an almost constant plateau for $s>0$ in Fig.\ref{fig:sm_bcd} (e). By entering the regime (b) ($w>v_1,v_2$) the singlet structure changes, resembling the situation Fig.\ref{fig:sm_bcd} (b). The spins at site 1 and site 2 are less correlated and $t_{\text{co}}$ is suppressed. In Fig.\ref{fig:sm_bcd} (e) for $0.25 < s < 0.75$ this leads to an drop of $t_{\text{co}}$, which is more pronounced the larger $w$ is compared to $v_1,v_2$. It is now possible to reach again the initial co-tunneling amplitude in the regime (a) by increasing the coupling $v_2$ at the bond 3-4. In regime (c) ($v_2 >w> v_1$), the initial singlet pairings are approximately restored and the co-tunneling $t_{\text{co}}$ is again enhanced as we see in Fig.\ref{fig:sm_bcd} (e) for $s>0.75$. Surprisingly, the described effect at bond 1-2 is reached by tuning only the couplings at the 2-3 and 3-4 bonds while $v_1$ at 1-2 is always constant (see Fig.\ref{fig:sm_bcd} (d)). However, the observation becomes clear by understanding how the spins are roughly paired in the groundstate and how this affects the probed \gls{gfz}. 
In Fig.\ref{fig:sm_bcd} (f), we normalized each curve by the initial value at $s=0$, to highlight this effect.
Our presented setup offers a qualitative signature, precise fine-tuning or knowledge of the tunnelings $v_1,v_2,w$, correlation strength $U$ or the lead-system coupling $C$ is not necessary as long as one probes roughly in our perturbative regime $\gamma \ll w,v_1,v_2$, $w,v_1,v_2 < U$.

%%%%%%%%%%%%%%%%%%%%%%%%%%%%%%%%%%%%%%%%%%%%%%%%%%%%%%%%%

\begin{thebibliography}{81}
\bibitem [{ref()}]{reference_sm}%
	\BibitemOpen
	\href@noop {} {\bibinfo {title} {See {S}upplemental material [url] for
			technical details, which includes refs. [51-54]}}\BibitemShut {NoStop}%
\end{thebibliography}
\end{document}